\documentclass[letterpaper,titlepage,12pt]{article}
\usepackage{hyperref}
\usepackage{amssymb,amsmath,amsfonts,bm}
\usepackage{epsfig}
\usepackage{times,cite,nicefrac}

\def\NN{{\cal N}}

\def\di{\text{d}}

\long\def\symbolfootnote[#1]#2{\begingroup%
\def\thefootnote{\fnsymbol{footnote}}\footnote[#1]{#2}\endgroup}
\newcommand{\oao}[2]{{#1\atopwithdelims[]#2}}
\newcommand{\oaop}[2]{{#1\atopwithdelims()#2}}

\def\llangle{ \langle \hskip-.9mm \langle}
\def\rrangle{ \rangle \hskip-.9mm  \rangle}

\newcommand{\tr}{\mathop{{\rm Tr}}}

\numberwithin{equation}{section}

\begin{document}

\begin{titlepage}

\rightline{\vbox{\small\hbox{0020308/RI}}}
\vskip 2.8cm

\centerline{\LARGE D-branes at multicritical points}

\vskip 1.6cm
\centerline{\bf Matthias R. Gaberdiel$^\clubsuit$, 
Dan Isra\"el$^\spadesuit$ and  
Eliezer Rabinovici$^\diamond$\symbolfootnote[1]{Email: \tt 
mrg@itp.phys.ethz.ch, israel@iap.fr, eliezer@vms.huji.ac.il 
}}
\vskip 0.5cm
\centerline{\sl $\clubsuit$ 
Institut f\"ur Theoretische Physik, \textsc{eth} Z\"urich, 
8093 Z\"urich, Switzerland
}
\centerline{\sl  $^\spadesuit$
\textsc{greco}, Institut d'Astrophysique de Paris,
98bis Bd Arago, 75014 Paris, France\symbolfootnote[2]{Unit\'e mixte 
de Recherche
7095, CNRS -- Universit\'e Pierre et Marie Curie}}
\centerline{\sl $^\diamond$
Racah Institute of Physics, The Hebrew University, Jerusalem 91904, Israel}

\vskip 1.4cm

\centerline{\bf Abstract} \vskip 0.5cm 

\noindent
The moduli space of  $c=1$ conformal field theories in two dimensions 
has a multicritical point, where a circle theory  is equivalent to an 
orbifold theory. We analyse all the conformal branes in both
descriptions of this theory, and find convincing evidence that the full brane spectrum
coincides. This shows  that the equivalence of the two
descriptions at this multicritical point extends to the boundary
sector. We also perform the analogous analysis for one of the
multicritical points of the ${\cal N}=1$ superconformal 
field theories at $c=\tfrac{3}{2}$. Again the brane spectra are
identical for both descriptions, however the identification  
is more subtle.

\vfill

\end{titlepage}
\tableofcontents

\section{Introduction}

Moduli spaces of two-dimensional conformal field theories
(\textsc{cft}s) often contain {\it multicritical points}, where
different lines (or hypersurfaces) of theories, generated by a set of
marginal operators, intersect. These points are quite interesting as
they admit several equivalent, but superficially different,
descriptions. More precisely, each formulation of the multicritical
point has the same set of primary fields (w.r.t.\ the chiral algebra
of the model), whose operator product expansions are given by the same
structure constants. However, the different formulations typically 
have different spacetime interpretations. For example, for the case of
the bosonic \textsc{cft} at $c=1$, the circle theory at twice the
self-dual radius is equivalent to the $\mathbb{Z}_2$ orbifold of the
self-dual circle theory \cite{Bardakci:1987ee,Ginsparg:1987eb}.

A natural question that arises in this context is whether these
seemingly different descriptions of the same bulk \textsc{cft} admit
the same set of boundary conditions and boundary operators when
defined on surfaces with boundaries, i.e.\ the same D-branes and open
string sectors in the string theory language. As the D-branes are
non-perturbative objects in space-time, they can be thought of as
refined probes compared to fundamental strings \cite{Shenker:1995xq}. 
They could therefore uncover unexpected distinctions between the 
different descriptions of the multicritical points. 

One approach to the construction of D-branes is in terms of boundary
states that lie in a suitable completion of the closed string
spectrum. These states are certain linear combinations of the
Ishibashi states \cite{Ishibashi:1988kg} that have to satisfy a
number of consistency conditions \cite{Cardy:1989ir,Cardy:1991tv}. This
construction can be performed purely in terms of the closed string
theory, and thus will be the same for the different descriptions of
a given multicritical point. However, {\it a priori}, it is not
guaranteed that only one set of consistent boundary states exists for
a given closed string theory, and thus the different descriptions of
a multicritical point may have different D-brane spectra.\footnote{For
D-branes that preserve the full rational symmetry of a rational
conformal field theory it was only recently shown that this cannot
happen \cite{KR}, but in the non-rational case (that shall concern us
here) the situation is unclear.} Indeed, each description of the
multicritical point has a specific space-time interpretation, and thus
comes with a certain set of `canonical' D-branes: for example, the
circle theory is described by an action and it implies that the 
theory must have Dirichlet and Neumann branes (since these can be
defined in terms of the action); similarly, the orbifold of the
circle theory contains always the orbifold invariant combinations of
the branes of the circle theory, etc. Since the various D-branes
of a theory must be mutually consistent (and since one can determine
the open string \mbox{moduli} from the open string spectra), these
`canonical' D-branes typically determine the full D-brane spectrum of
the bulk theory uniquely. It is then an interesting question to ask
whether these D-brane spectra agree for the different descriptions of
a multicritical point.

The multicritical points are also special since they have additional 
marginal operators (relative to the marginal operators that exist at a
generic point on each line) that allow one to move   along the
other branch. According to the general analysis of boundary 
deformations~\cite{Recknagel:1998ih} one may then expect that there is
a similar enhancement of the brane moduli space. However, in general
the analysis is more complicated since the brane moduli space may also
be bigger than what one would expect from the (preserved)
symmetries of the bulk theory \cite{Fredenhagen:2007rx}.
\smallskip

In this paper we shall analyse the complete D-brane spectra for the
different descriptions of various multicritical points. We shall
concentrate on those cases where we have full control of the bulk and
brane moduli spaces of the different theories. Besides the minimal
models (that do not admit marginal deformations) the only examples
known so far are free theories of bosons and fermions with central
charge and (local) superconformal symmetry \mbox{$(c=1$, $\NN=0)$}, 
\mbox{$(c=\tfrac{3}{2}$, $\NN=1)$} and  
\mbox{$(c = 3$, $\NN=2)$} 
\cite{Friedan,Gaberdiel:2001zq,Janik:2001hb,Gaberdiel:2004nv}. The
first two cases are more interesting, as both 
the bulk and brane moduli spaces are quite rich. In both cases, the
bulk moduli space has multicritical points (one for $\NN=0$ and five
for $\NN=1$), and the brane moduli space has components that are
described by discrete quotients of SU(2), that exists only for particular
discrete values of the compactification radius.  It was found
in~\cite{Friedan,Gaberdiel:2001zq,Janik:2001hb} that, on top of the
usual Dirichlet and Neumann branes, this accounts for a complete set
of boundary states. (For certain radii, the Dirichlet or
Neumann branes may be part of the continuous component.
Note also that these boundary conditions all define unitary open
string spectra; more general non-unitary boundary conditions are 
parametrised by elements in SL(2,$\mathbb{C}$).)

In this paper we work out the complete dictionary between the full brane
moduli spaces in the \mbox{$(c=1$, $\NN=0)$} model at the multicritical
point, where the circle and the orbifold lines
intersect~\cite{Ginsparg:1987eb}.\footnote{Previous work in this
direction has been done 
in~\cite{Recknagel:1998ih}. Following 
\cite{Gaberdiel:2001zq}, the relevant boundary states are much more
explicitly known, and our analysis can therefore be more complete.} 
In the \mbox{$(c=\tfrac{3}{2}$, $\NN=1)$} theory, all but one
multicritical point have rather uninteresting brane moduli spaces as
they do not contain many superconformal boundary states beyond the
usual Dirichlet and Neumann branes. We will therefore focus on the
multicritical point, where the circle theory and the
super-orbifold theory become equivalent~\cite{Dixon:1988ac}. 
For both multicritical points we shall 
find that the D-brane spectra of the two descriptions agree; in the
superconformal case however a number of intricate subtleties arise
that we explain in detail.
\smallskip

The plan of the paper is as follows. Section~\ref{cequalone} is
devoted to the analysis of the $c=1$ models, with a review of the
necessary material. In section~\ref{hatcequalone} we consider the
$\hat{c}=1$ ($c=\tfrac{3}{2}$) theories. Finally, section~4 contains our 
conclusions.

\section{The c=1 multicritical point}
\label{cequalone}

The moduli space of $c=1$ conformal field theories is well
known~\cite{Ginsparg:1987eb}. First of all there is the family of models
corresponding to a free real boson $X(z,\bar z)$ at radius $R$, the
{\it circle line}. This model has a U(1) symmetry that is enhanced to
an $\hat{\rm su}(2)_1 \times \hat{\rm su}(2)_1$ affine algebra
at the {\it self-dual radius} $R=\tfrac{1}{\sqrt{2}}$.\footnote{As
in most of the literature on this subject, we work in the units where
$\alpha'= 1/2$.} Secondly, the moduli space contains a line of
theories that are obtained by the $\mathbb{Z}_2$ orbifold
$\mathcal I~: X \mapsto -X$ from the circle theory; this family of
theories is usually called the {\it orbifold line}. Finally, the
moduli space contains isolated points obtained by orbifolding 
the circle theory at the self-dual radius by the
tetrahedral, octohedral and icosahedral discrete subgroups of SU(2).
They play no role in our discussion. (Branes for these
theories have been constructed in~\cite{Cappelli:2002wq}.)

The circle and orbifold lines intersect for $R_{\rm circ}=\sqrt{2}$
and $R_{\rm orb}=\tfrac{1}{\sqrt{2}}$. In fact, the circle theory at
twice the self-dual radius is also an orbifold (that is sometimes
denoted by SU(2)/$\mathcal{C}_2$), and it is not difficult to show
that  it is in fact equivalent to the above $\mathbb{Z}_2$ orbifold 
(that is sometimes denoted by SU(2)/$\mathcal{D}_1$)
\cite{Ginsparg:1987eb}.

\subsection{Boundary states of the circle line: a reminder}

We first review the construction of the boundary states on the circle
line following~\cite{Friedan,Gaberdiel:2001zq,Janik:2001hb}. For a
generic radius $R$, the  brane moduli space is rather simple: it
consists of standard Dirichlet and Neumann branes that preserve the
U(1) chiral symmetry of the model, each parametrised by a U(1) moduli
(the position of the brane and the Wilson line, respectively). In
addition there is a continuous family of branes that only couple to
the vacuum representation of the U(1) theory. 

For radii of the form $R=\tfrac{M}{N\sqrt{2}}$, with $M$ and $N$
coprime, the situation is more interesting \cite{Gaberdiel:2001zq}: in
addition to the Neumann and Dirichlet branes there are additional 
{\it conformal branes} preserving only the Virasoro algebra
that are parametrised by elements in
SU$(2)/\mathbb{Z}_M\times \mathbb{Z}_N$. Their origin can be
understood from two different points of view. First, one can think of
the theory at radius $R=\tfrac{M}{N\sqrt{2}}$ 
as a $\mathbb{Z}_M \times \mathbb{Z}_N$ orbifold of the 
self-dual circle theory, where $\mathbb{Z}_M$ ($\mathbb{Z}_N$) acts
via momentum (winding) shifts \cite{Tseng:2002ax}.\footnote{Alternatively, 
it can be viewed as the orbifold of the SU(2) level $k=1$ theory by 
$\mathcal{C}_M \times \tilde{\mathcal{C}}_N$, where   
the $\mathcal{C}_M$  ($\tilde{\mathcal{C}}_N$)
orbifold acts  vectorially (axially).}  By taking orbifold invariant
combinations, the conformal branes of the self-dual circle theory
(that are parametrised by SU(2)) then lead to the above moduli space
of conformal branes. 

Secondly, we can classify all conformal Ishibashi states, and then
construct the corresponding boundary states from them as in 
\cite{Gaberdiel:2001zq}.
To explain this in more detail we note that if the radius of the circle is 
$R=\tfrac{M}{N\sqrt{2}}$,
some of the $\hat{\rm u}(1)$ representations decompose into infinitely
many Virasoro representations. This happens whenever the 
conformal weight of a U(1) primary is of the form 
$\Delta=r^2$, $r \in \mathbb{N}/2$ (i.e.\ when the U(1) representation
has charge $q=\sqrt{2}r$), and the complete decomposition into 
Virasoro irreducible representations is of the form 
\begin{equation}
\mathcal{V}^{\mathfrak{u}_1}_{\sqrt{2} r}  = 
\bigoplus_{\ell=0}^\infty \mathcal{V}^{\mathfrak{Vir}}_{(r+\ell)^2}\ .
\end{equation}
According to Cardy's analysis~\cite{Cardy:1991tv} to each of these
representations one can associate an Ishibashi state that is a
solution of the gluing condition $T(z)=\bar T (\bar z)$  on the real
axis. In this way one obtains Virasoro Ishibashi states 
$|j,m,\bar m \rrangle$, where the quantum number $j \in \mathbb{N}/2$
denotes the irreducible Virasoro representation of conformal weight
$\Delta=j^2$, while $m$, $\bar m$ (with $j-m$ and 
$j-\bar m$ integer, $|m|, |\bar m| \leqslant j$) parametrise the 
$\hat{\rm u}(1)_{\rm L} \times \hat{\rm u}(1)_{\rm R}$
representations in which this Virasoro Ishibashi state appears 
\cite{Recknagel:1998ih,Gaberdiel:2001xm,Gaberdiel:2001zq}. Depending
on the specific choice of the radius, there are certain selection
rules which the $m$ and $\bar{m}$ have to satisfy.

Let us now specialise the discussion to the case $(M=2,N=1)$,  the
multicritical point. From the point of view of the circle line  
it corresponds to a compactification at radius $R=\sqrt{2}$, i.e.\
twice the self-dual radius. It is obtained as a $\mathbb{Z}_2$ winding   
shift orbifold of the self-dual point,
$\tilde X \sim \tilde X + \pi/\sqrt{2}$. The left and right conformal
weights of the  $\hat{\rm u}(1)\times \hat{\rm u}(1)$ primaries read
\begin{equation}\label{bosspec}
\Delta= \left( \frac{n}{4}+w \right)^2 \equiv m^2 \ , \qquad 
\bar{\Delta} =\left( \frac{n}{4}-w \right)^2 \equiv \bar{m}^2 \ , 
\end{equation}
where $n$ ($w$) denotes the momentum (winding) number. Note that the
theory possesses a pair of marginal operators with $(n=0,w=\pm 1)$; 
a certain linear combination of these states is the marginal operator
that allows one to move along the orbifold line. 

Let us discuss the boundary states of this theory. 
As for any radius, there are standard Dirichlet and Neumann branes. 
The Dirichlet branes involve the Dirichlet Ishibashi states 
$|(n,0)\rrangle$ built on momentum primaries with quantum numbers
$(n,0)$ , while the Neumann Ishibashi states $|(0,w)\rrangle$ arise in
the winding sectors $(0,w)$. In addition we have the  
Virasoro Ishibashi states $|j,m,\bar m \rrangle$ that come from the
degenerate $\hat{\rm u}(1)$ representations. At $R=\sqrt{2}$, the
relevant constraint on $m$ and $\bar{m}$ is simply 
\begin{equation}
m-\bar m \equiv 0 \mod 2\ . 
\label{momorb}
\end{equation}
We can think of this constraint as coming from an orbifold action,
as the circle theory at $R=\sqrt{2}$ is the $\mathcal{C}_2$ orbifold 
of the self-dual circle theory. In terms of the SU(2) description, the
orbifold acts as 
$g \mapsto \sigma_3 g \sigma_3 \equiv {\rm Ad}_{\sigma_3} g$. Thus the 
conformal branes of the circle theory at $R=\sqrt{2}$ are of the form
\cite{Gaberdiel:2001zq}
\begin{equation}
|\!|g\rrangle_{\rm circ} = 2^{1/4} \sum_{j,m,\bar m} 
\frac{1}{2} \Bigl[ D^j_{m,\bar m} (g) + 
D^j_{m,\bar m} \left({\rm Ad}_{\sigma_3}\, g \right)  \Bigr]
|j,m,\bar m\rrangle\ , 
\label{circbdystate}
\end{equation}
where $D^j_{m,\bar m} (g)$ is the SU(2) matrix element of $g$ in the 
$j$-representation. Obviously $g$ and ${\rm Ad}_{\sigma_3}\, g$
describe the same boundary state, and thus the moduli space of these
branes is SU(2)$/\mathbb{Z}_2$.\footnote{Note  that T-duality 
acts on the boundary states, and so on the brane moduli space, as 
$g \mapsto g\, i\sigma_1$. One 
obtains the moduli space of the T-dual theory, at radius 
$\tilde R = \nicefrac{1}{2\sqrt{2}}$.}

For off-diagonal group elements, i.e.\ group
elements of the form $g=i \sigma_1 e^{i\sigma_3 \phi}$, the resulting
brane is in fact a usual Neumann brane with Wilson line 
$\tilde x_0 =\phi/\sqrt{2}$ (that only involves the Ishibashi
states $|(0,w)\rrangle$). On the other hand, for diagonal group
elements $g= e^{i \sigma_3 \theta}$ with $\theta \in [0,2\pi)$, the
above brane is not fundamental, but rather describes the superposition
of two Dirichlet branes at opposite points on the circle. One way to
see this is to note that the condition on $m$ and $\bar{m}$
in~(\ref{bosspec}) to be half-integer implies that only
Ishibashi states with {\it even} $n$ appear in the boundary
state~(\ref{circbdystate}) --- the superposition of two Dirichlet
branes at opposite points on the circle projects out the odd momentum
states and only couples to even momentum states, see for example
(\ref{Dirichletbs}) below.  
The fact that these branes are not fundamental is also natural from a 
standard orbifold point of view. The branes (\ref{circbdystate})
are the usual {\it regular branes} of the orbifold
SU(2)/$\mathcal{C}_2$. As in other examples~\cite{Douglas:1996sw}
such branes can be resolved into {\it fractional branes} whenever
they sit on a fixed point of the orbifold. Since the group action is
$g\mapsto {\rm Ad}_{\sigma_3}\, g$, the fixed point set consists just
of the diagonal SU(2) matrices. If $g$ is diagonal, the corresponding
brane can be resolved into a pair of fundamental Dirichlet branes, 
whose boundary states include both even and odd momentum Ishibashi states
\begin{equation}
|\!|D;\theta\rrangle = 2^{-\nicefrac{3}{4}} \sum_{n \in \mathbb{Z}} 
e^{i \frac{n \theta}{2}} |(n,0)\rrangle \ .
\label{Dirichletbs}
\end{equation}
\begin{figure}[!ht]
\centering
 \includegraphics[width=80mm]{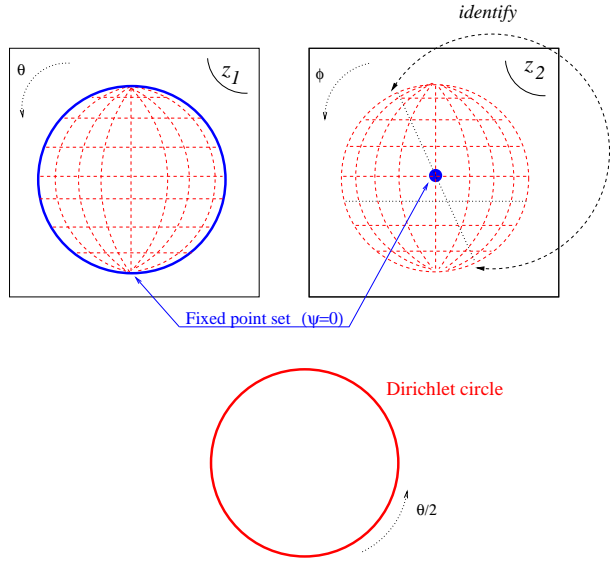} 
\label{circmod}
\caption{\it Two sections of the brane moduli space of the circle
theory, with the Dirichlet circle added. Antipodal  
points in the $z_2$-plane are identified.}
\end{figure}
To summarise, the single brane moduli space is made of the quotient
SU(2)/$\mathcal{C}_2$ with the circle $g= e^{i \sigma_3 \theta}$
removed, together with a disconnected $U(1)$ component that describes the
single Dirichlet brane moduli space.  It is convenient to
parametrise the SU(2) group elements in terms of Euler
angles\footnote{With this convention the Cartan-Weyl metric is  
$\di s^2 =  -\frac{1}{2} \tr (g^{-1} 
\di g\,  g^{-1} \di g) = \di \psi^2 +  \cos^2 \psi \, \di \theta^2 
+  \sin^2 \psi\,  \di \phi^2$.}
\begin{equation}
g = e^{ \frac{i\sigma_3}{2} (\theta+\phi)} 
e^{i \sigma_1 \psi} e^{ \frac{i\sigma_3}{2} (\theta-\phi)}\ .
\label{Eulang}
\end{equation}
Then the $\mathcal{C}_2$ orbifold acts as $\phi \mapsto \phi + \pi$. 
The fixed point set of this orbifold is $\psi=0$, i.e.\ the circle
spanned by $\theta \in [0,2\pi)$. Using the embedding of the
three-sphere in  $\mathbb{C}^2$ 
\begin{equation}
z_1 = \cos \psi \, e^{i\theta}\ , \qquad z_2 = \sin \psi\,  e^{i\phi}\ ,
\end{equation}
one gets the representation of the moduli space of
fig.~1.\footnote{The Neumann branes corresponds  
to the unit circle $z_1=0$, $|z_2|=1$, and are part
of the connected component of the moduli space of 
conformal branes.} The fixed point set of the
orbifold should be  removed from the one-brane moduli space. Instead
one has a separate component of the moduli space, isomorphic to
$U(1)$, describing a single  Dirichlet brane sitting on the circle at
$x=\theta/\sqrt{2}$ with $\theta \in [0,4\pi)$.  

Finally, it is interesting to notice that the Dirichlet branes are 
not sensitive to the underlying $SU(2)$ symmetry of the theory, as their 
moduli space describes a disconnected $U(1)$ component of the full
brane moduli space.

\paragraph{Comments on D-brane moduli space and target space geometry} 

One may ask whe\-ther there is any interesting relation between 
the moduli space of D-branes, and the target space geometry.
In particular, one may expect that the moduli space
of the D0-brane describes a `quantum version' of the target
space geometry.  However, an immediate problem with this idea 
is that it is a priori not clear how to characterise the D0-branes 
(among all conformal branes, say). Typically, the full moduli
space of conformal branes may contain different connected
components, and one may wonder whether all of them
lead to the same geometric interpretation, or whether
different brane probes may see `different geometries'. 

For example, in standard torus compactifications, the brane 
moduli space contains (typically) different connected components
corresponding to  different Dp-branes. These different
components of the moduli space lead in general to different 
geometries. However, the resulting geometries are all related by 
T-duality to one another (and so describe equivalent string 
backgrounds); this is simply a consequence of the fact that every 
Dp-brane can be transformed, by a suitable T-duality transformation, 
to  a D0-brane. Thus, in this case, every such component of the
brane moduli space can be thought of as a D0-brane moduli
space, and hence leads to a geometric interpretation in the 
above sense. The different geometries one obtains in this way
describe the different `geometrical interpretations' one may give
to the string theory in question. All of them describe, however,
equivalent conformal field theories. 

For the circle theory we have been considering so far, the
full D-brane moduli space is under control, and one may
test this idea further. For a generic radius, the full brane
moduli space contains three connected components:
the connected part of the moduli space of the Virasoro or conformal 
branes --- this excludes the Neumann and Dirichlet branes unless 
they are already part of it (which only happens is the radius is 
a multiple or a fraction of the self-dual radius);  the
moduli space of Dirichlet branes, and the moduli space of Neumann 
branes. In each case we should then consider the sigma model
whose target space is the brane moduli space in question.
In order to define such a sigma model one obviously needs to have a 
metric on the brane moduli space; this can be naturally obtained 
from the open string correlation functions. It is a priori not obvious
whether the resulting sigma-model is conformal; to make it conformal 
will sometimes require that certain $B$-fields are also switched on. 
For this to work the open string marginal operator correlation functions
in the appropriate sector should encode a $B$-field, just as they encode a
metric. Such a $B$-field is, for example, required for the case of
the moduli space of the conformal
branes which (for a `rational' radius)  is the
quotient $SU(2)/\mathbb{Z}_M \times \mathbb{Z}_N$; with the $B$-field
the resulting conformal sigma-model is then the
$\mathbb{Z}_M \times \mathbb{Z}_N$ orbifold of the $SU(2)$ \textsc{wzw}
model at level $k=1$. So the `geometry' that is associated to these
branes is $SU(2)/\mathbb{Z}_M \times \mathbb{Z}_N$.

The analysis for the Neumann or Dirichlet moduli spaces is simpler:
both are equivalent to a circle, which describes, on the nose, a conformal
sigma-model. The only difference is that the radius of the circle we obtain
in this manner is different for the Dirichlet and Neumann branes. However,
the two radii are related to one another by T-duality (as in the case of the 
Dp-branes mentioned above), and thus the resulting geometries
are equivalent in string theory. 

For the case at hand, the three brane moduli space components 
lead to a circle (for the case of the Neumann and Dirichlet branes),
and $SU(2)/\mathbb{Z}_M \times \mathbb{Z}_N$ (for the case
of the conformal branes). Superficially, the latter geometry is 
rather different from a circle; however, as a conformal field theory,
it is precisely equivalent to it again: the circle theory at radius
$R=(M/N) R_{\rm sd}$ is a 
$\mathbb{Z}_M \times \mathbb{Z}_N$ orbifold of the self-dual radius
theory, which in turn is equivalent to the level $k=1$ $SU(2)$ \textsc{wzw}
model! So even in this more complicated example, it seems that 
the different components of the brane moduli space lead to 
equivalent quantum geometries. It would be very interesting to 
analyse whether this observation generalises to other backgrounds 
as well.

\subsection{Boundary states of the orbifold}

The orbifold acts on the circle theory by the involution
$\mathcal I~: Y(z,\bar z) \mapsto - Y(z,\bar z)$, where in order 
to avoid confusion with the circle description we now 
denote the boson by $Y$.  The closed string
spectrum consists of the invariant states of the circle theory; in
addition we have two twisted sectors that are built on the two 
twist fields $\sigma^{y_0} (z,\bar z)$ (associated with the 
two fixed points at $y_0=0,\pi R$). These twist fields have 
conformal dimension $\Delta= \bar \Delta = 1/16$, and the oscillators
are half-integer moded in the twisted sector.

Let us now specialise to the self dual radius $R=1/\sqrt{2}$. As at
any radius we have the usual Dirichlet branes that are obtained by 
symmetrising the circle Dirichlet branes under the orbifold action
\begin{equation}
|\!|D;y\rrangle_{\rm orb } = 
2^{1/4} \sum_{n \in \mathbb{Z}} \cos (\sqrt{2} n y) \, 
|(n,0)\rrangle \ , 
\end{equation}
where $y \in  (0,\pi /\sqrt{2})$. These branes are fundamental,
except if they sit at one of the two fixed points 
$y_0=0,\pi/\sqrt{2}$. If this is the case, one can resolve them into
fractional Dirichlet branes that also involve the twisted sector
Ishibashi states (where $y_0=0,\pi/\sqrt{2}$) 
\begin{equation}
| D; y_0 \rrangle_{\rm tw} = 
e^{\sum_{r=1/2}^\infty
\frac{1}{r} b_{-r} \bar{b}_{-r}}\
\sigma^{y_0} |0 \rangle \ . 
\label{twistcard}
\end{equation}
The resulting {\it fractional branes} are then \cite{Oshikawa:1996dj}
\begin{equation}
|\!| D, y_0 ; \pm \rrangle_{\rm orb } = 
\tfrac{1}{2}\, |\!|D;y_0 \rrangle_{\rm orb} 
\pm 2^{-\nicefrac{1}{4}} 
| D; y_0 \rrangle_{\rm tw} \ ,
\label{fracDir}
\end{equation}
where $y_0=0,\pi/\sqrt{2}$. 
The situation is similar for the Neumann branes. 
The orbifold invariant combination of the Neumann branes are 
\begin{equation}
|\!|N;\tilde{y}\rrangle_{\rm orb } = 
2^{1/4} \sum_{w \in \mathbb{Z}} \cos (\sqrt{2} w \tilde{y}) \, 
|(0,w)\rrangle \ . 
\end{equation}
The appropriate combination of Neumann Ishibashi states from the
twisted sector reads  
\begin{equation}
| N; \tilde{y}_0  \rrangle_{\rm tw} = 
\frac{1}{\sqrt{2}} \, e^{-\sum_{r=1/2}^{\infty}
\frac{1}{r} b_{-r} \bar{b}_{-r}}\
\left[ \sigma^{0} + e^{i\sqrt{2} \tilde{y}_0} \, 
\sigma^{\frac{\pi}{\sqrt{2}}}\right] |0 \rangle \ , 
\label{Ntwistcard}
\end{equation}
where $\tilde{y}_0=0,\tfrac{\pi}{\sqrt{2}}$. As for the Dirichlet
branes, the full Neumann fractional brane boundary states are then 
\begin{equation}
|\!| N, \tilde{y}_0 ; \pm \rrangle = 
\tfrac{1}{2}\, |\!|N;\tilde{y}_0 \rrangle_{\rm orb} 
\pm 2^{-\nicefrac{1}{4}} 
| N; \tilde{y}_0  \rrangle_{\rm tw} \ .
\label{fracNeu}
\end{equation}

These are the well-known standard constructions, but one can also
obtain the additional conformal boundary states for this
theory. Indeed, the orbifold can be thought of as the SU(2) orbifold
by $\mathcal{D}_1$. 
\begin{figure}[!ht]
\centering
 \includegraphics[width=80mm]{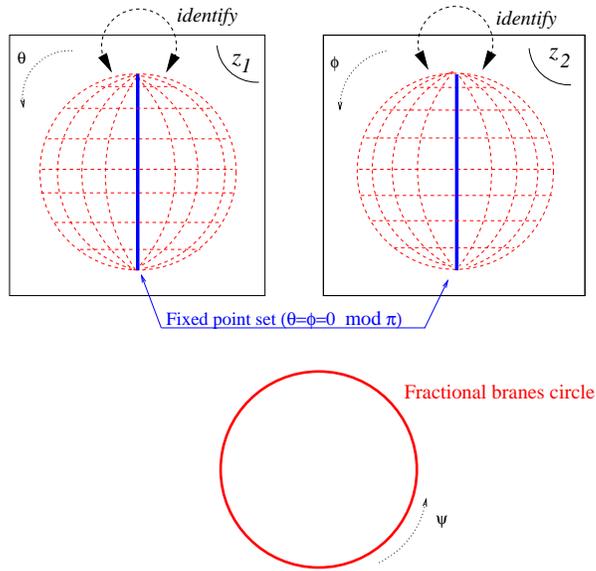} 
\label{orbmod}
\caption{\it Two sections of the brane moduli space of the orbifold
  theory, with the fractional branes circle added. }  
\end{figure}
This orbifold acts on SU(2) group elements as 
$g \mapsto \sigma_1 g \sigma_1$. Indeed, the operator $\mathcal I$
acts on $\hat{\rm u}(1) \times \hat{\rm u}(1)$ primaries as 
$(m,\bar m) \mapsto (-m,-\bar m)$, and hence maps the 
Virasoro Ishibashi states as 
\begin{equation}
|j,m,\bar m\rrangle  \mapsto |j,-m,-\bar m\rrangle \ . 
\end{equation}
This is then reproduced by $g \mapsto \sigma_1 g \sigma_1$ since (see
(\ref{matel}) for the definition of these matrix elements) 
\begin{equation}
D^j_{-m -\bar m} (g) = (-)^{m-\bar m } D^j_{m \bar m} (g^\star) = 
D^j_{m \bar m} (\sigma_3 g^\star \sigma_3 ) = 
D^j_{m \bar m} (\sigma_1 g \sigma_1)\ .
\end{equation}
Thus the orbifold invariant boundary states are simply 
\begin{equation}
|\!|g\rrangle_{\rm orb} = 2^{1/4} \, 
\sum_{j,m,\bar m} \frac{1}{2} \Bigl[
D^j_{m,\bar m} \left(g \right) + 
D^j_{m,\bar m} \left({\rm Ad}_{\sigma_1} \, g \right) \Bigr]\,
|j,m,\bar m\rrangle\ .  
\label{orbbdystate}
\end{equation}
The branes are fundamental unless $g$ belongs to the fixed point set
of the orbifold, i.e.\  unless $g$ is a matrix of the form 
$g =  e^{i\sigma_1 \psi}$. It contains the particular cases 
$\psi=0,\pi$ and $\psi =  \pi/2, 3\pi/2$. The former  
corresponds to the Dirichlet brane at the two fixed points,
while the latter are the Neumann branes at the T-dual fixed points --- see
the discussion above. We postpone  the construction of the fractional 
branes for generic $\psi$ to the next subsection.  
For the moment we only note that, in terms of the Euler angles
parametrisation of SU(2), see~(\ref{Eulang}), the orbifold acts as  
$\theta \mapsto -\theta$ and $\phi \mapsto -\phi$, and the fixed point
set corresponds to $\theta=\phi=0$. Using the same embedding in
$\mathbb{C}^2$ one gets the moduli space  
depicted in fig.~2.

\subsection{Mapping of  moduli spaces and resolution of the orbifold
  branes}  

In the previous subsection we have obtained the full brane moduli 
space of the circle and the orbifold theories at the multicritical
point. For the circle theory, there is a continuous family of
conformal branes parametrised by $SU(2)/\mathcal{C}_2$, with a circle
removed corresponding to the fixed point set of the orbifold. This
circle is replaced by a disconnected $S^1$ component corresponding to
the moduli space of a single Dirichlet brane. For the orbifold theory
we also have a continuous family of conformal branes parametrised by
$SU(2)/\mathcal{D}_1$ (the orbifold is different, although
isomorphic), again with a circle removed corresponding to the fixed
point set of the orbifold. We expect on general grounds that the
corresponding branes can be resolved into two fractional branes; for
special points (that correspond either to standard Dirichlet and
or Neumann branes) we have already described this above, and we shall
construct the resolution for the other branes below. It is
interesting to note (as was already observed in
\cite{Recknagel:1998ih}) that the family of resolved branes
interpolates between fractional Dirichlet and fractional Neumann
branes. 

Our strategy will be to {\it assume} that the two brane moduli spaces are 
equivalent. We can then use the known resolution of the fixed point branes
of the circle theory --- these are pairs of Dirichlet branes at opposite points 
on the circle, that are resolved into single Dirichlet branes --- in order to 
determine what the resolution of the fixed point branes 
in the orbifold theory should be. As we shall see, the resulting picture 
satisfies a number of non-trivial consistency conditions, thus suggesting
that our identification has been correct.

\paragraph{Relating the generic conformal branes} 

We start by identifying the generic conformal branes of 
both theories that are parametrised by SU$(2)/\mathbb{Z}_2$. As we
have explained above, these branes contain the standard Neumann branes
of the circle theory, as well as the standard (non-fixed point)
Neumann and Dirichlet branes of the orbifold.  The correspondence must relate
the two different $\mathbb{Z}_2$ actions into one another, and one 
finds that this is achieved by the rotation matrix 
of Euler angles $(\psi=\pi/4;\theta=\pi/2;\phi=0)$, i.e.\ the map 
\begin{equation}
\rho~: \ g_{\rm circ} \mapsto  g_{\rm orb} = \rho(g_{\rm circ}) = 
h\, g_{\rm circ} \,  h^{-1} \quad \text{with} \qquad h = \frac{i}{\sqrt{2}} 
\left(
\begin{array}{cc} 1 & 1 \\ 1 & -1 \end{array} \right) \ , 
\label{circorbmap}
\end{equation}
which has the property 
$\rho (\sigma_3 g \sigma_3) = \sigma_1 \rho(g)\sigma_1$. Thus we have
the dictionary
\begin{equation}
|\!| g \rrangle_{\rm circ} \Longleftrightarrow 
|\!|\rho(g) \rrangle_{\rm orb} \ . 
\end{equation}
Under this map, the pairs of antipodal Dirichlet branes 
in the circle theory (i.e.\ the group elements 
$g=e^{i\sigma_3 \theta}$) are mapped to orbifold branes with
\begin{equation}
g_\textsc{d} = \left(\begin{array}{cc}  \cos \theta & i\sin \theta \\ 
i\sin \theta &  \cos \theta \end{array}\right)\ ,
\label{dircircle}
\end{equation}
corresponding to the fixed point set of the orbifold. These are
neither Dirichlet (diagonal $g$)  nor Neumann (off-diagonal $g$)
boundary states, except for particular values of $\theta$. For  
$\theta =\{0;\pi\}$ (the `Dirichlet points') one gets a regular
Dirichlet brane sitting at one of the two orbifold fixed points, while
for $\theta = \{\tfrac{\pi}{2};\tfrac{3\pi}{2} \}$ (the `Neumann
points') one gets a regular Neumann brane at a T-dual fixed point.   

As another special case, the Neumann branes of the circle theory are
mapped to orbifold branes with  
\begin{equation}
g_\textsc{n} = \left(\begin{array}{cc} i \sin \psi & -\cos \psi \\ 
\cos \psi & -i \sin \psi \end{array}\right)\ .
\label{neucircle}
\end{equation}
These are fundamental branes for any $\psi$, as are the
corresponding Neumann branes in the circle theory.

\paragraph{Resolution of the orbifold branes} 

Before we can discuss the dictionary between the fractional branes, we
first need to understand the resolution of the fixed point branes in
the orbifold theory. We expect on general grounds that the resolved
branes will involve Ishibashi states from the twisted sector. We
therefore need to understand first what Virasoro Ishibashi states
appear in the twisted sector of the orbifold. 

The twisted sector $\hat{\rm u}(1)$ representation of the orbifold is
not an irreducible Virasoro representation; for example, the state
$b_{-1/2}\sigma^{y_0}$ is a Virasoro highest weight state. On the
other hand, the conformal dimensions that appear 
($\Delta = \tfrac{1}{16} + \tfrac{N}{2}$ with $N\in\mathbb{N}$) all
correspond to non-degenerate Virasoro representations. Thus one can
determine the decomposition based on the comparison of
characters. The chiral twisted character of a single fixed point is 
\begin{equation}
\sqrt{ \frac{\eta (\tau)}{\vartheta_4 (\tau)} } = 
\frac{\sqrt{\vartheta_2 (\tau) \vartheta_3 (\tau) }}{\sqrt{2} \eta
  (\tau)}   
= \frac{\vartheta_2 (\tau/2) }{2 \eta (\tau)} 
= \sum_{\ell=0}^{\infty} 
\frac{q^{\frac{1}{4}(\ell+\nicefrac{1}{2})^2}}{\eta (\tau)}\ ,
\end{equation}
where we have used standard $\vartheta$-function identities, see
appendix~\ref{appconv}. The characters that appear on the right hand
side are all characters of irreducible Virasoro representations. By
the standard argument we therefore obtain Virasoro Ishibashi states
from the twisted sector labelled by 
$$
|(\ell+\nicefrac{1}{2})^2/4;y_0 \rrangle_{\rm Vir} \ ,
$$
where $y_0=0,\tfrac{\pi}{\sqrt{2}}$ labels the two fixed points. 
Note that these Ishibashi states are in one-to-one correspondence with
the Dirichlet Ishibashi states $|(n,0)\rrangle$ of the circle theory
corresponding to odd $n$. (This is in fact guaranteed by the
equivalence of the two closed string spectra.)
We fix their normalisation by requiring that 
\begin{equation}
|D; y_0 \rrangle_{\rm tw} = \sum_{\ell=0}^{\infty} 
|(\ell+\nicefrac{1}{2})^2/4;y_0\rrangle_{\rm Vir} \ . 
\label{fracdecomp}
\end{equation}
For the Neumann twisted sector Ishibashi states, we have to take care
of the extra minus sign in the coherent state, see
eqn.~(\ref{Ntwistcard}). It follows that if the Virasoro highest
weight vector involves an odd number of oscillators,  the
corresponding Ishibashi state appears with a relative minus sign.
A Virasoro highest weight vector of conformal weight 
$\Delta = 1/16 + \ell(\ell+1)/4$ is made out of an  
even number of oscillators provided that 
$\ell(\ell+1)/4 \in \mathbb{Z}$, 
i.e.\ for $\ell\equiv 0$ or $1 \mod 4$. Thus it follows that 
\begin{multline}
|N; \tilde{y}_0 \rrangle_{\rm tw} 
= \frac{1}{\sqrt{2}}\sum_{\ell=0}^{\infty} 
(-)^{\frac{\ell (\ell+1)}{2}}
\left[ |(\ell+\nicefrac{1}{2})^2/4;0 \rrangle_{\rm Vir} 
+ e^{i\sqrt{2} \tilde{y}_0} 
|(\ell+\nicefrac{1}{2})^2/4;\nicefrac{\pi}{\sqrt{2}} 
\rrangle_{\rm Vir} \right] \ .
\label{fracdecompN}
\end{multline}

Because of the symmetries of the problem there 
is some arbitrariness in the mapping between the (fractional) 
branes. We choose to identify the pair of Dirichlet branes at
$x=\{0;\pi\sqrt{2}\}$ with the unresolved orbifold Dirichlet brane 
sitting at the fixed point  \mbox{$y=0$}, consistently with the 
map~(\ref{circorbmap}). It implies that the single Dirichlet brane at
$x=0$ is mapped for instance to 
the Dirichlet fractional brane~(\ref{fracDir}) with positive sign;
with this convention the single Dirichlet brane at $x=\pi\sqrt{2}$
corresponds then to the fractional Dirichlet brane~(\ref{fracDir})
with negative sign. By comparing the boundary state
~(\ref{fracdecomp}) with the odd-momenta coupling
of~(\ref{Dirichletbs}) it follows that we identify 
\begin{equation}
|D;(2\ell+1,0)\rrangle + 
|D;(-2\ell-1,0)\rrangle = 
|(\ell+\nicefrac{1}{2})^2/4;y_0=0 \rrangle_{\rm Vir}  \ . 
\label{dic1}
\end{equation} 
The pair of Dirichlet branes at 
$x=\{\nicefrac{\pi}{\sqrt{2}};\nicefrac{3\pi}{\sqrt{2}}\}$ in the
circle theory should then be identified with  
the unresolved orbifold Dirichlet brane sitting at the fixed point
$y=\nicefrac{\pi}{\sqrt{2}}$, using~(\ref{circorbmap}).  This leads to
the identification
\begin{equation}
i (-)^\ell \Big[ |D;(2\ell+1,0)\rrangle - 
|D;(-2\ell-1,0)\rrangle \Big] = 
|(\ell+\nicefrac{1}{2})^2/4;y_0=\nicefrac{\pi}{\sqrt{2}} 
\rrangle_{\rm   Vir}  \ . 
\label{dic2}
\end{equation} 
These two linear relations (\ref{dic1}, \ref{dic2}) now fix the
dictionary between the twisted sector Ishibashi states of both  
theories completely. Since we know already how to resolve the fixed
point branes of the circle theory (these are just standard single Dirichlet
branes) we can now make an ansatz for the resolved branes of the
orbifold theory, using these relations. This leads to 
\begin{eqnarray}
|\!|  e^{i\sigma_1 \psi};\pm \rrangle_{\rm orb} 
& = & \tfrac{1}{2} \, |\!| g= e^{i\sigma_1 \psi}
                \rrangle_{\rm orb} \nonumber \\
& & \pm \, 2^{-\nicefrac{1}{4}} \sum_{\ell=0}^{\infty} \Bigl\{ 
\cos ((\ell+\tfrac{1}{2})\psi) \,  |(\ell+\nicefrac{1}{2})^2/4;0 \rrangle_{\rm Vir}  
\\
& & \qquad \qquad \qquad + (-)^\ell 
\sin ((\ell+\tfrac{1}{2})\psi) \,  
|(\ell+\nicefrac{1}{2})^2/4;\nicefrac{\pi}{\sqrt{2}} 
\rrangle_{\rm Vir}\Bigr\} \ . \nonumber
\label{contfrac}
\end{eqnarray}
As a consistency check of this analysis we can now verify that the
branes at $\psi=\tfrac{\pi}{2}$ and $\psi=\tfrac{3\pi}{2}$  
correspond to the Neumann branes with Wilson lines $\tilde{y}_0=0$ and
$\tilde{y}_0=\tfrac{\pi}{\sqrt{2}}$, respectively. Indeed, we find
that, for instance,  
\begin{multline}
|\!| e^{i\sigma_1 \pi/2};\pm \rrangle_{\rm orb}  =   
\tfrac{1}{2}\, |\!|  e^{i\sigma_1 \pi/2}\rrangle_{\rm orb} \\
 \pm\, 2^{-\nicefrac{3}{4}} 
\sum_{\ell=0}^{\infty}  (-)^{\frac{\ell (\ell+1)}{2}} 
\Bigl\{ |(\ell+\nicefrac{1}{2})^2/4;0 \rrangle_{\rm Vir}  
 +   |(\ell+\nicefrac{1}{2})^2/4;\nicefrac{\pi}{\sqrt{2}} 
\rrangle_{\rm Vir}\Bigr\}\ , 
\end{multline}
which is exactly the same expression as the Neumann fractional brane
boundary state  with Wilson line $\tilde{y}_0 = 0$,  see
eqn.~(\ref{fracNeu}) together with (\ref{fracdecompN}). The analysis
for $\psi = 3\pi/2$ is essentially identical.
\medskip

In summary we have thus constructed the resolution of the 
general conformal fixed point branes of the orbifold theory by 
assuming that the branes of the circle and the orbifold theory
are equivalent. Our ansatz has passed the fairly non-trivial consistency 
check that it reproduces correctly the known resolution of both Dirichlet and
Neumann branes of the orbifold theory. We regard this as  convincing 
evidence that the proposed resolution, as well as the identification of the
brane moduli spaces, is indeed correct. 

Incidentally, our analysis also
implies that all of these resolved branes of the orbifold theory actually 
preserve a $\hat{\rm u}(1)$ symmetry, not just the Virasoro
symmetry (because the corresponding branes of the circle theory do); 
the relevant chiral currents are 
$(J,\bar J) = (\cos 2\sqrt{2} X_L (z), \cos 2\sqrt{2} X_R(\bar{z}))$. 
Unfortunately, it is however rather difficult to verify this
directly.

\section{The \^c=1 moduli space}
\label{hatcequalone}

Next we want to consider the $\mathcal{N}=(1,1)$ superconformal field
theories with $\hat c = \tfrac{2c}{3} = 1$. As for its bosonic
counterpart  considered in the previous section, there is a continuous
moduli space of such theories, with additional isolated points. All
lines of theories can be obtained as orbifolds of the theory of a free
boson and a free Majorana fermion. Since there are a number of
different orbifolds one may consider (involving in particular the
fermion number), the moduli space is quite rich and contains five
different lines of theories (circle, orbifold, 
super-affine, orbifold-prime and super-orbifold),  intersecting
pairwise at five different multicritical points (we refer the reader
to~\cite{Dixon:1988ac} for  details). In order to obtain a modular
invariant partition function we have to chose a
\textsc{gso}-projection; for the circle theory we shall consider the
0B-type \textsc{gso}-projection in the following.\footnote{At
$\hat{c}=1$ only the 0A- and 0B-type \textsc{gso}-projections are
possible. Obviously, after \textsc{gso}-projection, the
theory is not an $\mathcal{N}=(1,1)$ superconformal field theory any
more since the supercurrents are projected out.}

Among the multicritical points, four are such that none of the
non-trivial super-$\hat{\rm u}(1)$ representations that appear in the
spectrum are reducible with respect to the $\mathcal{N}=1$
super-Virasoro (SVir) algebra; at these multicritical points the brane
spectrum will therefore be rather small, and will in particular not
contain any component related to SU(2) besides the identity component that is common 
to all non-rational radii. However, this is not the case
for the multicritical point where the circle line at $R=1$ intersects the
super-orbifold line also at $R= 1$. In that case the spectrum contains
many degenerate SVir representations, and hence the brane moduli space
is fairly rich. This is not accidental as both theories are
(isomorphic) $\mathbb{Z}_2$-orbifolds of the super-affine
$\hat{\rm so}(3)_1 \times \hat{\rm so}(3)_1$ theory ,  
i.e.\ the theory of three left- and three right-moving Majorana-Weyl
fermions with identical boundary conditions on the torus. Since this
is the most interesting case, we shall restrict ourselves to this
multicritical point in the following.\footnote{This
is actually the only multicritical point among the five for which  the
superconformal, rather than the \textsc{gso}-projected, theories
are identical \cite{Wendland:2004pp}.} 

We shall be interested in the branes that preserve the $\mathcal{N}=1$
superconformal symmetry, i.e.\ that satisfy the gluing condition
\begin{equation}
\left( L_n - \bar{L}_{-n} \right) |\!|B;\eta \rrangle = 0\ ,  \qquad 
\left( G_r +i\eta\, \bar{G}_{-r} \right) |\!|B;\eta \rrangle = 0\ ,
\label{glucond}
\end{equation}
where $\eta=\pm 1$ labels the two possible superconformal gluing
conditions.  We will consider the two cases separately as they give
rise to different sets of branes and have subtle differences regarding
the dictionary of the D-branes.

\subsection{Branes of the circle line}
\label{supercircle}
We begin with the analysis of the free boson and fermion theory, i.e.\
the tensor product of the $c=1$ circle line (a free boson at radius
$R$) with an Ising model, i.e.\ a pair of Majorana-Weyl worldsheet 
fermions $\chi,\, \bar \chi$.  This is the
theory that was studied in \cite{Gaberdiel:2001zq}.

As before, we first have to find all SVir Ishibashi states, i.e.\ we
have to analyse which SVir representations are degenerate. 
In the Neveu-Schwartz (\textsc{ns}) sector a SVir representation is
degenerate if the highest weight has conformal dimension
$\Delta = \nicefrac{r^2}{2}$ with $r$ integer. The first null vector
is then at level $r +\nicefrac{1}{2}$, which has therefore the 
opposite worldsheet fermion number parity compared to the primary
state.  A super-$\hat{\rm u}(1)$ representation with  
charge $r$ decomposes then into irreducible SVir representations as
\begin{equation}
\mathcal{V}^{\mathfrak{u}_1,\, \textsc{ns}}_{r} 
= \sum_{\ell = 0}^{\infty} \mathcal{V}^{\mathfrak{SVir}\, ,\, 
\textsc{ns}}_{(r+\ell)^2/2} \ .
\label{decomprep}
\end{equation}
In this decomposition the SVir representations have (chiral)
worldsheet fermion number $F= \ell \mod 2$. In the Ramond (\textsc{r})
sector the SVir representations  are degenerate for 
$\Delta = \nicefrac{r^2}{2}+\nicefrac{1}{16}$ with 
$r \in \mathbb{N}+\nicefrac{1}{2}$ (the null vector is also at level
$r +\nicefrac{1}{2}$). Super $\hat{\rm u}(1)$ primaries with  
such conformal weights (in the \textsc{nsns} and \textsc{rr} sectors)
occur whenever the compactification radius is rational. We consider
the case $R=1$ below, relevant for the analysis of the
circle/super-orbifold multicritical point. The left and right
conformal dimensions of the $\hat{\rm u}(1)$ primaries then read
\begin{equation}
\Delta= \frac12 \left( \frac{n}{2}+w \right)^2 
+\frac{a^2}{16}\equiv \frac{m^2}{2} + \frac{a^2}{16}\ , 
\qquad \bar{\Delta} =\frac12 \left( \frac{n}{2}-w \right)^2
+ \frac{a^2}{16} \equiv \frac{\bar{m}^2}{2} + \frac{a^2}{16} \ , 
\label{confweightsup}
\end{equation}
where $a=0$ ($a=1$) in the \textsc{ns} (\textsc{r}) sector.

\paragraph{Boundary states in the NSNS sector}

Let us start with the \textsc{nsns} sector Ishibashi states. The
super-$\hat{u}(1)$ representations are reducible with respect to the
SVir algebra if the momentum $n$ is even. From the decomposition of
these representations into irreducible SVir $\times$  SVir
representations of dimension 
\mbox{$\Delta= \bar \Delta  =\nicefrac{j^2}{2}$},  
one obtains a set of Ishibashi states 
$|j,m,\bar m;\eta\rrangle_\textsc{ns}$, where $j\in\mathbb{N}$, $j-m$,
$j-\bar{m}$ are integer and $|m|,|\bar{m}|\leqslant j$. The label $\eta$
denotes the parameter in the gluing condition~(\ref{glucond}). The
superconformal branes in the \textsc{nsns} sector, constructed out of
this set of Ishibashi states, then read~\cite{Gaberdiel:2001zq} 
\begin{equation}
|\!|g;\eta\rrangle_\textsc{ns} =  \frac{1}{\sqrt{2}} \,
\sum_{j \in \mathbb{N},\, m,\bar m }
\Bigl[D^j_{m \bar m} \left( g\right) 
+ D^j_{m \bar m} \left({\rm Ad}_{\sigma_3}\,  g\right) \Bigr]\, 
|j,m,\bar m; \eta\rrangle_\textsc{ns}\ .
\label{superconfcircns}
\end{equation}
The sum in square brackets imposes the constraint that $m-\bar{m}$ is
even (since the winding number $w$ is integer); the constraint that
the momentum number $n$ is even is automatically satisfied since $j$
(and hence $m$ and $\bar{m}$) is integer. The  boundary state is
invariant under the diagonal  \textsc{gso}-projection for both
choices of $\eta$. Indeed, the \mbox{SVir $\times$ SVir}
representation with \mbox{$\Delta=\bar \Delta = j^2 /2$} is such   
that $j=m+\ell=\bar m + \bar \ell$. Since $m-\bar m$  is even, so is
$\ell - \bar \ell $, and it follows from the discussion below
eqn~(\ref{decomprep}) that all Ishibashi states are
\textsc{gso}-invariant.  

As before, the boundary state corresponding
to $g$ and to ${\rm Ad}_{\sigma_3}\,  g$ are the same. Furthermore,
branes parametrised by $g$ and $-g$ are identical, as 
$D^j_{m \bar m} (-g) = (-)^{2j} D^j_{m \bar m} (g)$. Thus the 
moduli space of \textsc{nsns} sector boundary states is 
SO$(3)/\mathbb{Z}_2$. 

For off-diagonal SU(2) group elements (i.e.\ group elements of the
form $g=i\sigma_1 e^{i\sigma_3 \phi}$), one gets an ordinary
\textsc{nsns} sector Neumann boundary state, with couplings to all
winding states  (as $m=-\bar m = w$ is not constrained). On the other 
hand, diagonal group elements ($g= e^{i\sigma_3 \theta}$) describe a
pair of Dirichlet branes since only even momentum states ($n$ even)
contribute.  The single Dirichlet branes occur as the resolution of
these branes; their boundary state is given by 
\begin{equation}
|\!|D;x;\eta\rrangle_\textsc{ns} 
= \frac{1}{\sqrt{2}}\sum_{n\in \mathbb{Z}} 
e^{i n x} | (n,0)\rrangle \otimes |D;\eta\rrangle_\textsc{ns}\ , 
\label{dirsupercirc}
\end{equation}
where $|D;\eta\rrangle_\textsc{ns}$ is the \textsc{nsns} Dirichlet
fermion Ishibashi state solving 
\begin{equation}
\Bigl( \chi_r -i\eta \bar{\chi}_{-r}\Bigr) 
|D;\eta \rrangle_\textsc{ns} =0\ . 
\label{DIshi}
\end{equation}
The parameter $x$ denotes the position of the brane with 
$x=\theta/2$ and $\theta\in [0,4\pi)$. 

\paragraph{Boundary states in the RR sector} 

In the \textsc{rr} sector the discussion is more subtle because of the 
fermionic zero modes. First we see from~(\ref{confweightsup}) that the
super-$\hat{\rm u}(1)$ representations are degenerate with respect to the
SVir algebra for odd momentum $n$. Analogously to what we have done
above, we thus get SVir Ishibashi states 
$|j,m,\bar m;\eta\rrangle_\textsc{r}$ where now $(j,m,\bar m)$ are
half-integer moded. The  boundary states then read
\begin{equation}
|\!|g;\eta\rrangle_\textsc{r} =  \frac{1}{\sqrt{2}}\, 
\sum_{j \in \mathbb{N}+\tfrac{1}{2},\, m,\bar m} 
\Bigl[ D^j_{m \bar m} \left( g \right) +
D^j_{m \bar m} \left({\rm Ad}_{\sigma_3}\,  g \right) \Bigr]
|j,m,\bar m; \eta\rrangle_\textsc{r}\ .
\label{superconfcircR}
\end{equation}
As before the boundary states associated to $g$ and 
${\rm  Ad}_{\sigma_3}\,  g$ are identical, but now the boundary state
corresponding to $g$ and $-g$ have opposite sign. The projection in
the square bracket guarantees, as before, that only terms with 
$m-\bar m$ even contribute.

The analysis of the \textsc{gso}-projection is slightly subtle. Using
the conventions of \cite{Gaberdiel:2001zq}, the superconformal
Ishibashi states satisfy 
\begin{equation}
(-)^{F+\bar F} |j,m,\bar m;\eta\rrangle_\textsc{r}  = 
- \eta (-)^{m - \bar m} |j,m,\bar m;\eta\rrangle_\textsc{r} \ ,
\end{equation}
while the fermionic Dirichlet and Neumann Ishibashi states obey
\begin{equation}\label{GSOnd}
(-)^{F+\bar F} |D;\eta\rrangle_\textsc{r} =  
-\eta \, |D;\eta\rrangle_\textsc{r} \ , \qquad
(-)^{F+\bar F} |N;\eta\rrangle_\textsc{r} =  
\eta \, |N;\eta\rrangle_\textsc{r} \ .
\end{equation}
Here $|D;\eta\rrangle_\textsc{r}$ is characterized by the gluing
condition (\ref{DIshi}), while the gluing condition for 
$|N;\eta\rrangle_\textsc{r}$ has the opposite sign, 
$(\chi_r +i\eta \bar{\chi}_{-r}) |N;\eta \rrangle =0$. 
Note that this is consistent with the fact that Dirichlet Ishibashi
states appear with $m=\bar m$, while for Neumann Ishibashi states we
always have $m=-\bar m$; in the \textsc{r}-sector $m$ and $\bar m$ are
half-integer, and hence $m-\bar m$ is odd. 

As before, the \textsc{rr}-sector boundary states
(\ref{superconfcircR}) are fundamental unless $g$ is diagonal. For
diagonal $g$, we have again a pair of Dirichlet branes which can be
resolved; the resolved single Dirichlet brane boundary state
\begin{equation}
|\!|D;x;\eta\rrangle_\textsc{r} = 
\frac{1}{\sqrt{2}}\sum_{n\in \mathbb{Z}} e^{i n x} 
| (n,0)\rrangle \otimes |D;\eta\rrangle_\textsc{r}
\label{dirsupercircR}
\end{equation}
then couples to even and odd momenta $n$, whereas
(\ref{superconfcircR}) only involves terms with odd $n$ (since $m$ and
$\bar{m}$ are half-integer). 

The full boundary states are finally obtained by putting 
\textsc{nsns}- and \textsc{rr}-sector boundary states together. Their
structure depends on the choice of $\eta$ in (\ref{glucond}). For
$\eta=-1$,  all SVir Ishibashi states are \textsc{gso}-invariant, and we get 
\begin{equation}
|\!|g;\eta=-1\rrangle = \frac{1}{\sqrt{2}} \Bigl( 
|\!|g;\eta=-1\rrangle_\textsc{ns}
+i |\!|g;\eta=-1\rrangle_\textsc{r} \Bigr) \ . 
\label{dirra}
\end{equation}
The brane with $-g$ is the antibrane of that with $g$, but we have the
identification that $g$ and ${\rm Ad}_{\sigma_3} g$ describe the same
D-brane. The moduli space is therefore SU$(2)/\mathbb{Z}_2$. These
branes are fundamental unless $g$ is diagonal; for diagonal $g$ we can 
resolve the brane into a Dirichlet brane anti-brane pair sitting at
opposite points on the circle. The full boundary state of a single
Dirichlet brane is of the form 
\begin{equation}
|\!|D;x;- \rrangle = \frac{1}{\sqrt{2}} 
\Bigl( |\!|D; x;-\rrangle_\textsc{ns} 
+ i |\!|D; x;-\rrangle_\textsc{r} \rangle  \Bigr) \ . 
\end{equation}
On the other hand, for $g$ off-diagonal, the brane (\ref{dirra})
describes a Neumann brane that does not involve a \textsc{rr}
component --- these branes are the analogues of the non-\textsc{bps}
branes that were first studied by Sen~\cite{Sen:1998sm}. This is in
agreement with the fact that for $\eta=-1$ the \textsc{rr}-sector
Neumann Ishibashi states are \textsc{gso}-odd --- see (\ref{GSOnd}).  
\medskip

For $\eta=+1$, all \textsc{rr} SVir Ishibashi states
$|j,m,\bar{m}\rrangle_{\textsc{rr}}$ are \textsc{gso}-odd. Thus for 
$\eta=+1$ the boundary states $|\!|g;\eta=+1\rrangle_\textsc{ns}$ are
consistent by themselves. As discussed above, the corresponding moduli
space is SO$(3)/\mathbb{Z}_2$. The branes are fundamental provided
that $g$ is neither diagonal nor off-diagonal: the branes with $g$
diagonal describe the superposition of two non-\textsc{bps} Dirichlet
branes at opposite points on the circle (that only involve
\textsc{nsns}-sector Ishibashi states). For $g$ off-diagonal, on the
other hand, the above $\mathbb{Z}_2$ (that inverts the signs of the
off-diagonal matrix elements of $g$) acts trivially on SO$(3)$ and the
branes get resolved into a superposition of a Neumann brane
anti-brane pair, i.e.\ in branes with opposite \textsc{rr}
charge. (For $\eta=+1$, the \textsc{rr}-sector Neumann 
Ishibashi states are \textsc{gso}-invariant.)

\subsection{Super-affine line } 

Although the multicritical point  we are interested in lies on the
intersection of the circle and the super-orbifold line, it is useful
to discuss first the branes of the super-affine line, as it contains
the point of maximally extended symmetry in the $\hat c=1$ moduli
space. Indeed for $R=1$ the super-affine theory is identical to a
theory of three Majorana fermions with identical boundary conditions,
realising an  $\hat{\rm so}(3)_1 \times \hat{\rm so}(3)_1$  
Ka\v c-Moody algebra with $\mathcal{N}=(1,1)$ supersymmetry. 

The super-affine theory can be obtained from  the circle theory as the
orbifold by the symmetry $\mathcal S\, : \ s_\textsc{x} (-)^{F_{st}}$,  
where $s_\textsc{x}$ is the shift symmetry
$X\mapsto X+\pi R$ while $F_{st}$ denotes the left-moving 
{\it spacetime  fermion number}. Thus $(-)^{F_{st}}$ acts as $+1$
($-1$) on the  
\textsc{nsns} (\textsc{rr}) sector. The full torus partition function of the
super-affine theory is 
\begin{equation}
\mathcal Z (\tau) = \frac{1}{2} \sum_{\gamma, \delta \in \mathbb{Z}_2}
\sum_{n,w\in \mathbb{Z}} (-)^{\delta n }  
\frac{q^{\frac{1}{8} \left(\frac{n}{R}+R(2w+\gamma) 
\right)^2}}{\eta (\tau )}
\frac{\bar{q}^{\frac{1}{8} 
\left(\frac{n}{R}-R(2w+\gamma) \right)^2}}{\bar \eta (\bar \tau)} 
\frac{1}{2} \sum_{a,b \in \mathbb{Z}_2} (-)^{a\delta+b\gamma 
+\gamma\delta} \left|
\frac{\vartheta 
\left[ {a \atop b} \right]}{\eta}\right|  . 
\label{torussuperaff}
\end{equation}
Note that (as is usual for $(-)^{F_{st}}$ orbifolds), the 
\textsc{gso}-projection is reversed in the twisted sector
($\gamma=1$). Furthermore, the spacetime fermion number parity is
opposite in the twisted sector, i.e.\ it acts as $-1$ in the twisted 
\textsc{nsns} sector and as $+1$ in the twisted \textsc{rr} sector.

\paragraph{Boundary states of the super-affine theory at R=1}
Let us now construct the branes of the super-affine theory at $R=1$,
starting from the circle theory at $R=1$. In the 
untwisted \textsc{nsns} sector, the orbifold by $\mathcal{S}$ has the
effect of projecting onto even-momentum states, which implies that all
the superconformal Ishibashi states 
$|j,m,\bar m ;\eta\rrangle_\textsc{ns}$ are or\-bi\-fold-in\-va\-riant
(as they all have  $n=m+\bar m$ even by construction). On the other
hand, the additional Dirichlet Ishibashi states with $n$ odd are
projected out. Thus all superconformal Ishibashi states in the
untwisted sector \textsc{nsns} sector are of the form 
$|j,m,\bar m ;\eta\rrangle_\textsc{ns}$ with $j$ integer and 
$m-\bar m$ even.

In the untwisted \textsc{rr} sector,  the \textsc{rr} ground states
are odd  under $(-)^{F_{st}}$, and thus only odd momentum states
survive. Again all the superconformal Ishibashi states 
$|j,m,\bar m ;\eta \rrangle_\textsc{r}$ are invariant under the
orbifold projection (since $n=m+\bar m$ is now odd), while the even
momentum states in (\ref{dirsupercircR}) are again projected out. The
same also applies to the Neumann Ishibashi states (for which
$n=0$). Thus all superconformal Ishibashi states in the untwisted 
\textsc{rr} sector are of the form 
$|j,m,\bar m ;\eta\rrangle_\textsc{ns}$ with $j$ half-integer and 
$m-\bar m$ even.

In the $\mathcal S$-twisted sector of the super-affine theory, the
winding number $w$ is half-integer (see eqn.~(\ref{torussuperaff}) for
$\gamma=1$). In the twisted \textsc{nsns}-sector, we therefore get
superconformal Ishibashi states labelled by 
$|j,m,\bar m ;\eta \rrangle_{\textsc{ns};\, \rm tw}$, where $j$ is
still integer, but now $m-\bar m$ 
is odd. As mentioned above, the ground state of the twisted
\textsc{nsns} sector is now \textsc{gso}-odd, and therefore these
Ishibashi states survive the  \textsc{gso}-projection. They also
survive the orbifold projection by $\mathcal{S}$ since $n=m+\bar m$ is
now odd, but the twisted  \textsc{nsns} sector is also odd under
$(-)^{F_{st}}$. Taken together with the Ishibashi states of the
untwisted \textsc{nsns} sector we therefore have SVir Ishibashi
states associated to $(j,m,\bar m)$, where $j$ is integer, but without
any constraints on $m-\bar m$.  

The analysis for the Ishibashi states from the twisted \textsc{rr}
sector is similar. We find superconformal Ishibashi states labelled by
$|j,m,\bar m ;\eta \rrangle_{\textsc{r};\, \rm tw}$, where $j$ is
half-integer, but now $m-\bar m$ is odd.
Since the GSO-projection in the twisted sector is opposite to what one
had before, they are \textsc{gso}-invariant for $\eta=-1$ (and  
\textsc{gso}-odd for $\eta=+1$). Furthermore they are orbifold
invariant since $n=m+\bar m$ is now even as is the 
$(-)^{F_{st}}$ parity of the twisted \textsc{rr}. Together with the 
superconformal Ishibashi states of the untwisted \textsc{rr}
sector we therefore get SVir Ishibashi states associated to 
 $(j,m,\bar m)$, where $j$ is half-integer, but without
any constraints on $m-\bar m$. These states are only
\textsc{gso}-invariant for $\eta=-1$. 

It is now straightforward to construct the boundary states of the
super-affine theory. In the \textsc{nsns} the boundary states are
simply given by 
\begin{equation}
|\!|g\rrangle_{\textsc{ns};\, {\rm s-a}} =
 \sum_{j\in \mathbb{N}, m,\bar m} 
D^j_{m \bar m} \left(   g \right) |j,m,\bar m; \eta\rrangle_\textsc{ns}\ ,
\label{superaffns}
\end{equation}
where $j\in\mathbb{N}$, $m,\, \bar m \in \mathbb{Z}$ with
$|m|, |\bar m| <j$, but with no further restrictions on $m,\bar m$. In 
the \textsc{rr} sector we have instead 
\begin{equation}
|\!|g\rrangle_{\textsc{r};\, {\rm s-a}} =
 \sum_{j\in \mathbb{N}+\frac{1}{2}, m,\bar m} 
D^j_{m \bar m} \left(   g \right) |j,m,\bar m; \eta\rrangle_\textsc{r}\ ,
\label{superaffr}
\end{equation}
where $j, m$ and $\bar{m}$ are now half-integer with
$|m|, |\bar m| \leqslant j$, but again with no further restrictions on 
$m,\bar m$. The \textsc{rr} sector boundary state is only
\textsc{gso}-invariant if $\eta=-1$; otherwise it is \textsc{gso}-odd.  

Putting the boundary states together, their structure depends on the
choice of $\eta$. If $\eta=+1$, the \textsc{nsns} boundary state is
already consistent by itself. Since only integer $j$ appear in the
sum, the moduli space of these branes is precisely SO$(3)$. The
annulus amplitude in the open string channel is 
\begin{equation}
\mathcal A  (t)  = \left(\frac{\vartheta_3 (it)}{\eta (it)} \right)^{3/2}\, , 
\end{equation} 
i.e.\ it corresponds to three free fermions with identical boundary
conditions. Since the sum over $m$ and $\bar m$ is not further
restricted, all of these boundary states are fundamental. In
particular, these boundary states describe a single Dirichlet brane for 
diagonal $g$, and a single Neumann brane for off-diagonal $g$. 

For $\eta=-1$, we can add the \textsc{rr} sector
contribution, and  
the full boundary state is 
\begin{equation}
|\!|g;\eta=-1\rrangle_{{\rm s-a}} = \frac{1}{\sqrt{2}} \Bigl(
|\!|g;\eta=-1\rrangle_{\textsc{ns};\, {\rm s-a}} + i 
|\!|g;\eta=-1\rrangle_{\textsc{r};\, {\rm s-a}} \Bigr) \ .
\end{equation}
Since there are no identifications, the moduli space of these branes
is precisely SU$(2)$. The open string channel annulus amplitude now
reads  
\begin{equation}
 \mathcal A (t)  = \frac{1}{2} \left[ 
\left(\frac{\vartheta_3(it)}{\eta(it)}\right)^{3/2}\!\!\! 
+ \ \left(\frac{\vartheta_4 (it)}{\eta(it)}\right)^{3/2}\right]\ .
\end{equation}
This describes precisely the \textsc{gso}-projected spectrum of three
free fermions (with identical boundary conditions). 

Compared to the circle theory, the moduli space of branes is much
simpler since it does not involve any non-trivial identifications. The
super-affine theory at $R=1$ is therefore the natural analogue of the
bosonic circle theory at the self-dual radius. As such, it is the
natural starting point from which the branes of the other theories
should be constructed.

\paragraph{The circle line from the super-affine line} 

In order to check this idea we should, in particular, be able to
reproduce the branes of the circle theory (at $R=1$) starting from the
simple branes of the super-affine theory at $R=1$. The circle theory
at $R=1$ is the orbifold of the super-affine theory at $R=1$ by
the winding shift orbifold
$s_{\tilde{\textsc{x}}} : \tilde X \mapsto \tilde X + \pi/R$.  In the
untwisted sector, this simply removes the sectors for which the
winding number if half-integer. The twisted sectors of the 
$s_{\tilde{\textsc{x}}}$-orbifold describe odd-momentum states
in the \textsc{nsns} sector, and even-momentum states in the 
\textsc{rr} sector, and one therefore obtains indeed the circle theory. 

The winding shift orbifold acts on the superconformal Ishibashi states
of the super-affine theory at $R=1$ as $(-)^{m-\bar m}$. In terms of
the boundary states that are labelled by $g$ this action becomes
\begin{equation}
s_{\tilde{\textsc{x}}} \, |\!|g\rrangle_{{\rm s-a}} = 
|\!|\sigma_3\, g\,\sigma_3 \rrangle_{{\rm s-a}} \ . 
\end{equation}
The unresolved branes of the circle theory are then simply the sums
$$
\frac{1}{\sqrt{2}} \Bigl( |\!|g\rrangle_{{\rm s-a}} + 
|\!|{\rm Ad}_{\sigma_3}\, g\rrangle_{{\rm s-a}} \Bigr) \ .
$$
These are precisely the superconformal boundary states of the circle
theory at $R=1$, see eqn.~(\ref{dirra}) for $\eta=-1$ and 
eqn.~(\ref{superconfcircns}) for $\eta=+1$. Fixed points appear for 
$g$ diagonal, and for the case of $\eta=+1$ also for $g$
off-diagonal. (Recall that for $\eta=+1$, the group elements live in
SO(3)=SU(2)$/\mathbb{Z}_2$.) 
The corresponding branes then need to be resolved; this
involves adding in $s_{\tilde{\textsc{x}}}$-twisted sector states. 
For both values of $\eta$ we have $s_{\tilde{\textsc{x}}}$-twisted
sector Dirichlet Ishibashi states in the \textsc{nsns} sector with odd 
momentum. For $\eta=-1$ we have in addition 
$s_{\tilde{\textsc{x}}}$-twisted sector Dirichlet Ishibashi states in
the \textsc{rr} sector with even momentum, while for 
$\eta=+1$ there are $s_{\tilde{\textsc{x}}}$-twisted sector Neumann
Ishibashi states in the \textsc{rr} sector with all winding numbers. 
Adding in these twisted sector components then allows us to resolve
the fixed point branes: for diagonal $g$ we obtain the single
Dirichlet branes of the circle theory. (For $\eta=-1$ the resolution
involves adding in \textsc{nsns} and \textsc{rr} components, while for
$\eta=+1$ only twisted \textsc{nsns} sector states are used.) Finally,
for $\eta=+1$ and $g$ off-diagonal, we obtain single Neumann branes
by using the twisted \textsc{rr} sector Neumann Ishibashi states.

\subsection{Super-orbifold line}

After this brief interlude we can now return to the super-orbifold
line that intersects the circle line at $R=1$. The super-orbifold
theory can be obtained as an orbifold of the super-affine line
by the inversion $\mathcal I~:\ Y \mapsto -Y$, 
$\chi \mapsto -\chi$, $\bar \chi \mapsto -\bar \chi$. The
branes are obtained from the super-affine branes in a very similar
fashion as for the $c=1$ theories discussed previously in
sec.~\ref{cequalone}.

We restrict the discussion to the super-orbifold theory at radius
$R=1$. By the usual argument, we obtain invariant boundary states by
adding together branes and their image under the orbifold action; this
leads to 
\begin{equation}
|\!|g\rrangle_{{\rm s-o}}  
= \frac{1}{\sqrt{2}} \Bigl( 
|\!|g\rrangle_{{\rm s-a}} + 
|\!|\sigma_1 \,g \,\sigma_1 \rrangle_{{\rm s-a}} 
\Bigr)\ .
\label{regbraneso}
\end{equation}
This describes a fundamental brane unless $g$ is a fixed point. For
$\eta=-1$ where $g\in {\rm SU}(2)$, the only fixed points are the
group elements of the form $g=  e^{i\sigma_1 \psi}$, i.e.\
\begin{equation}\label{sof1}
g = \left( \begin{matrix}
\cos\psi & i \sin\psi \\ i \sin\psi & \cos\psi
\end{matrix} \right) \ . 
\end{equation}
For $\eta=+1$
when $g\in {\rm SO}(3)$, we have additional fixed points 
when $\sigma_1 \, g \, \sigma_1 = - g$, i.e.\ for 
\begin{equation}\label{sof2}
\hat{g} = \left( \begin{matrix}
i\cos\theta & - \sin\theta \\  \sin\theta & - i \cos\theta
\end{matrix} \right) \ . 
\end{equation}
Both fixed point sets contain standard Dirichlet branes
($\psi, \theta=\{0;\pi\}$) and standard Neumann 
branes ($\psi, \theta=\{\pi/2;3\pi/2\}$).

\paragraph{Twisted sectors and fractional branes} In order to resolve
these fixed points we will need to add Ishibashi states from the
twisted sector of the $\mathcal{I}$-orbifold. In the twisted sector
the bosons are half-integer moded, while the fermions are integer
moded in the \textsc{ns} sector, and half-integer moded in the 
\textsc{r} sector. There are again two fixed points, and at each of
them we can decompose the chiral character into irreducible
representations of the SVir algebra. For the \textsc{ns} sector we
find 
\begin{equation}
\left(  \frac{\vartheta_2}{2\vartheta_4}\right)^{1/2} 
= \frac{\vartheta_2 \sqrt{\vartheta_3}}{2 \eta^{3/2}}
= \frac{1}{2}\sum_{n \in \mathbb{Z}} 
\frac{q^{\frac{1}{2}(n+\frac{1}{2})^2}}{\eta} 
\left( \frac{\vartheta_3}{\eta}\right)^{1/2} 
= \sum_{n = 0}^{\infty} \chi^{\mathfrak{SVir;\, 
\textsc{ns}}}_{ \frac{(n+1/2)^2}{2}}\ .
\label{svirnsdecomp}
\end{equation}
Each term in the sum over $n$ corresponds to the character of a
(non-degenerate) irreducible SVir representation. Thus we get two 
sets of superconformal Ishibashi states 
$|(n+1/2)^2/2,y_0;\eta\rrangle_{\textsc{ns}, {\rm SVir}}$, where 
$y_0=0$ and $y_0=\pi/2$. (Note that in the super-affine theory, $y$
and $y+\pi$ are identified, and hence $y_0=\pi/2$ is a fixed point of
the inversion.) Since these are \textsc{ns} SVir 
Ishibashi states, there are no zero modes for the supercurrents, and
the Ishibashi states are \textsc{gso}-invariant for both choices of 
$\eta$. We choose their normalisation such that the Dirichlet
Ishibashi states at these fixed points are simply
\begin{equation}\label{Dfix}
|D;y_0\rrangle_{\textsc{ns}; {\rm tw}} = \sum_{n=0}^{\infty}
|(n+1/2)^2/2,y_0;\eta\rrangle_{\textsc{ns}, {\rm SVir}} \ . 
\end{equation}
We observe that these Ishibashi states are in one-to-one
correspondence with the odd-momentum \textsc{nsns} sector Dirichlet
Ishibashi states of the circle theory (that appear in the twisted
sector of the $s_{\tilde{\textsc{x}}}$ orbifold). 

\noindent In each twisted \textsc{r} sector (i.e.\ for $y_0=0$ and 
$y_0=\frac{\pi}{2}$) we obtain instead
\begin{equation}
\left(  \frac{\vartheta_3}{\vartheta_4}\right)^{1/2} 
= \frac{1}{\sqrt 2} \frac{\vartheta_3 
\sqrt{\vartheta_2}}{ \eta^{3/2}}
= \sum_{n \in \mathbb{Z}} \frac{q^{\frac{n^2}{2}}}{\eta} 
\left( \frac{\vartheta_2}{2\eta}\right)^{1/2} 
= \sum_{n = 0}^{\infty} \chi^{\mathfrak{SVir;\, 
\textsc{r}}}_{ \nicefrac{n^2}{2}+\nicefrac{1}{16}}\ ,
\end{equation}
which we have again decomposed into a sum of (non-degenerate) irreducible
SVir representations. Note that the \textsc{r} representation with
$n=0$ has a single ground state (since it is annihilated by $G_0$),
whereas the representations with $n\neq 0$ have two ground states that
are related to one another by the action of $G_0$. We therefore get,
for both values of $\eta$, exactly one \textsc{rr} sector Ishibashi
state for $n=0$ since the zero-mode condition is now trivial! The two 
Ishibashi states corresponding to $y_0=0$ and $y_0=\tfrac{\pi}{2}$
have opposite \textsc{gso}-parity since the one with $y_0=0$ comes
from the untwisted sector of the $\mathcal{S}$-orbifold, while the one
at $y_0=\tfrac{\pi}{2}$ comes from the $\mathcal{S}$-twisted
sector. Thus only one of these two Ishibashi states survives the 
\textsc{gso}-projection; as we shall see later on, in the conventions
of this paper it is the one with $y_0=\tfrac{\pi}{2}$.

For $n\neq 0$, on the other hand, there are four ground states (each 
chiral representation has two ground states) and we get exactly two 
Ishibashi states $|n^2/2,y_0;\eta \rrangle_{\textsc{r}}$ for each value
of $\eta$. However the two Ishibashi states for given $\eta$ have
opposite \textsc{gso}-parity, and thus only one of them survives the 
\textsc{gso}-projection. Thus for each value of $\eta$ we have one
Ishibashi state $|n^2/2,y_0;\eta \rrangle_{\textsc{r}}$, where
$n\geq 1$ and where $y_0$ can take the two values $y_0=0$ and 
$y_0=\frac{\pi}{2}$. These are in one-to-one correspondence with the
even-(nonzero)-momentum \textsc{rr} sector Dirichlet Ishibashi states
of the circle theory for $\eta=-1$, or the non-zero-winding Neumann
Ishibashi states of the  circle theory for $\eta=+1$. (These are the
Ishibashi states of the circle theory that appear in the twisted
sector of the $s_{\tilde{\textsc{x}}}$-orbifold.) Obviously this  
had to be the case since the two closed string theories are
equivalent.

\subsection{The brane moduli space at the multicritical point} 

We are now in the position to establish the dictionary between the
superconformal branes of the circle and super-affine theories at
$R=1$. As discussed above the circle theory can be obtained as a
$\mathcal C_2$ orbifold of the super-affine theory at $R=1$, while
the super-orbifold theory corresponds to a $\mathcal D_1$
orbifold of the same theory. The situation is therefore very similar
to the bosonic case discussed in section~2. However, as we shall see
below, there are new subtleties associated with the \textsc{gso}
projection.   

\subsubsection{Identification of the unresolved branes} 

The identification between the unresolved superconformal branes is
as in the bosonic $c=1$ model. The two superconformal branes  
moduli spaces are related by the same rotation~(\ref{circorbmap}),
leading to the identification  
\begin{equation}\label{giden}
|\!| g;\eta \rrangle_{\rm circ} \Longleftrightarrow 
|\!|\rho(g) ;\eta\rrangle_{\rm s-o}\ . 
\end{equation}
This works for both values of $\eta$. Here $|\!| g;\eta \rrangle$ is
a \textsc{nsns} boundary state if $\eta=+1$; for $\eta=-1$, the branes
involve \textsc{nsns} as well as \textsc{rr} components. 

\subsubsection{Identification of the resolved branes} 
We are now left with the problem of identifying the fractional branes
of the super-orbifold theory (involving the twisted sectors) with the
corresponding boundary states of the circle theory. The analysis
depends crucially on the sign of the parameter $\eta$ which labels the
superconformal boundary conditions. We therefore consider each case
separately.

Our strategy will be to {\it assume} that the branes of the circle theory
are in one-to-one correspondence with those of the super-orbifold theory.
This will allow us to determine the twisted sector resolutions in 
the super-orbifold theory, by translating the known resolutions from the 
circle theory. As we shall see, the resulting picture is compatible with
the intrinsic constraints of the super-orbifold theory; this is a 
non-trivial consistency condition on the ansatz, and therefore
suggests that the assumed one-to-one correspondence of the branes is
indeed correct.

\paragraph{Correspondence of fractional branes for  $\bm{\eta=+1}$}
Let us start from the  circle theory branes analysed in
subsection~\ref{supercircle}. Before resolution these branes only have 
an \textsc{nsns} sector and are parametrised by
SO(3)$/\mathbb{Z}_2$. There are two circles of fixed points
corresponding to diagonal and off-diagonal matrices. 

The diagonal matrices describe two Dirichlet branes at opposite points
on the circle, and are resolved using odd-momenta \textsc{nsns}
Dirichlet Ishibashi states. Under  the map~(\ref{circorbmap}), these
fixed point branes correspond to the family of branes in the
super-orbifold theory that are described by $g$ as
in~(\ref{sof1}). From the super-orbifold point of view, this set of
branes contains one `Dirichlet point', namely $g=\mathbb{I}$ (for
$\psi=0$).\footnote{Note that unlike the bosonic case, there is only 
{\it one independent} Dirichlet point in the family of boundary 
states~(\ref{sof1}) since in SO(3) $g\cong-g$.} The corresponding
boundary state~(\ref{regbraneso}) is 
\begin{equation}
|\!| \mathbb{I} \rrangle = \sqrt{2}\sum_{m=-\infty}^\infty 
\sum_{j=|m|}^\infty |j\, m \, m ;\eta=+1\rrangle_\textsc{ns}
= \sqrt{2}\sum_{m=-\infty}^\infty |(2m,0)\rrangle \otimes 
          |D;\eta=+1 \rrangle_\textsc{ns} \ .
\end{equation}
We associate this boundary state with a regular Dirichlet brane
sitting at the orbifold fixed point $y=0$. 

This regular brane is resolved using the \textsc{nsns} twisted sector
Ishibashi states that come from the corresponding orbifold fixed point
(i.e.\ from $y=0$). The relevant combination is described by
eqn.~(\ref{Dfix}) with $y=0$. In the twisted \textsc{nsns} sector, the
fermions  $\chi,\bar\chi$  are periodic, so the fermionic ground
states belong to a two-dimensional representation  of the 
zero-mode algebra. However, only one linear combination is 
orbifold invariant, and it is the one that satisfies a Dirichlet boundary 
condition for $\eta=+1$, i.e.\ it is a solution of 
\begin{equation}
(\chi_0-i\eta \bar{\chi}_0 ) 
|0\rangle_{\textsc{ns};\mathcal{I}{\rm -tw.}} = 0 \quad 
\hbox{with $\eta=+1$.}
\label{diriground}
\end{equation} 
This state is then also \textsc{gso}-invariant.

The family of super-orbifold fixed point branes~(\ref{sof1}) also
contains  a `Neumann point' for $\psi=\pi/2$. It is resolved with a
Neumann twisted \textsc{nsns} sector Ishibashi state. Somewhat
surprisingly, this Ishibashi state is {\it only} associated with the
fixed point $y=\pi/2$. In fact, as we saw above, only the ground state
satisfying the Dirichlet boundary condition eqn.~(\ref{diriground}) is
orbifold- and \textsc{gso}-invariant in the twisted sector associated
with the fixed point  $y=0$. However, the fixed point $y=\pi/2$ is
`twisted' both with respect to the inversion
$\mathcal I$ and the super-affine symmetry $\mathcal{S}$, and hence
has the reversed \textsc{gso} and orbifold projection relative to $y=0$. 
Thus only the fixed point at $y=\pi/2$ supports a Neumann Ishibashi state 
(with $\eta=+1$) in the twisted \textsc{nsns} sector.
\medskip

\begin{figure}[!ht]
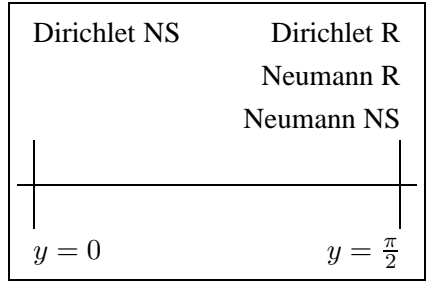

\centering
\fbox{\small \begin{tabular}{lcr}
Dirichlet NS & & Dirichlet R\\
&& Neumann R\\
&& Neumann NS \\
\vline &&  \vline \\
\hline 
\vline && \vline \\
$y=0$ && $y=\tfrac{\pi}{2}$ 
\end{tabular}}
\caption{\it Twisted sector branes for $\eta=+1$ in the super-orbifold
theory. }   
\label{etaplusfig}
\end{figure}

As we mentioned above, the circle theory for $\eta=+1$ has a second
set of fixed points that correspond to off-diagonal matrices. 
They correspond in the circle theory to \textsc{nsns} sector 
Neumann branes, which are resolved using \textsc{rr} Neumann 
Ishibashi states. In the super-orbifold theory, they are mapped to the
family of branes~(\ref{sof2}). This set of branes contains again a
Dirichlet point, this time corresponding to $g=i\sigma_3$ 
\begin{equation}
|\!| i\sigma_3  \rrangle = 
 \sqrt{2}\sum_{m=-\infty}^\infty (-)^m 
|(2m,0)\rrangle \otimes |D;\eta=+1 \rrangle_\textsc{r} \, ,
\end{equation}
which is naturally associated to the fixed point $y=\pi/2$. Translating
the circle theory resolution, we therefore have to resolve
this brane with twisted \textsc{rr} sector Ishibashi states.
In the twisted \textsc{rr} sector of the super-orbifold, the fermion has a
unique ground state (the fermions are half-integer moded), that can be
either \textsc{gso}-odd or \textsc{gso}-even. Given our conventions,
the correspondence suggests that the fixed point $y=\pi/2$ gives a
\textsc{gso}-even twisted \textsc{rr} ground state (so that the 
Dirichlet brane can be resolved by a fixed point contribution from the
`same' fixed point). This then implies that the \textsc{rr} ground
state at $y=0$ is \textsc{gso}-odd. 
Thus we only have $\hat{\mathfrak{u}}(1)$-preserving
Ishibashi states from the twisted \textsc{rr} sector at $y=\pi/2$, 
and the Ishibashi states from this sector must resolve
both the Dirichlet brane for $\theta=0\mod \pi$, as well as 
the Neumann brane for  $\theta=\pi/2 \mod \pi$. These results are
summarised in fig.~\ref{etaplusfig}.   

An interesting feature of this analysis is that a regular Dirichlet
brane (which has only a contribution from the \textsc{nsns} sector)
is resolved by a twisted sector \textsc{nsns} Ishibashi state at the fixed   
point $y=0$, while at the other fixed point $y=\pi/2$ it is resolved
using a twisted sector \textsc{rr} Ishibashi state. Another
somewhat surprising feature is that the Neumann branes get only
resolved by twisted sector contributions from one end, namely from the
fixed point at $y=\pi/2$. These properties are probably a consequence
of the somewhat unusual orbifold that defines the super-orbifold 
theory.

\paragraph{A tale of two etas} 
We shall below carry out a similar analysis for the superconformal 
gluing conditions with $\eta=-1$. Before delving into the details of this,
there is an important issue we need to explain. In the previous paragraph
(i.e.\ for  $\eta=+1$), the boundary state with $g=\mathbb{I}$ was   
identified with a Dirichlet brane sitting at the fixed point $y=0$ in
the super-orbifold theory. One may then suspect that the identification
between this group element and the specific fixed point will also hold
for the other gluing condition (i.e.\ $\eta=-1$), but as we shall now explain,
this is not correct.

In order to see this, let us consider the annulus amplitude involving
the two \textsc{nsns} boundary states, one associated with $g=\mathbb{I}$ 
for $\eta=-1$, and one with $g=i\sigma_3$ for $\eta=+1$
\begin{equation}
\mathcal{A}_{-+}=\, _\textsc{ns} \llangle \eta=-1;g=\mathbb{I} |\!| 
e^{-\pi t H_\textsc{c}} |\!|g=i\sigma_3 ;\eta=+1\rrangle_\textsc{ns} \ .
\label{crossannulus}
\end{equation}
In the closed string channel, this amplitude belongs to the
$\widetilde{\textsc{ns}}$ sector, i.e.\  it leads to a \textsc{nsns} character
with the insertion of $(-)^F$. The superconformal Ishibashi states are 
normalised such that\footnote{The phase in front of the right-hand-side is understood
as follows. The Ishibashi states 
$|j\, m\, m ;\eta\rrangle$ are built on an \textsc{ns} SVir.\  
primary of conformal weight $\Delta  =\nicefrac{j^2}{2}$, coming from
the decomposition of a reducible representation of lowest weight
$\Delta=\nicefrac{m^2}{2}$ (with $|m|\leqslant j$). If $j=|m|$, the
SVir primary state has even left-moving fermion number (since it is
identified with a super-$\hat{\mathfrak{u}}(1)$ primary).  For
$j=|m|+1$, the SVir irreducible representation is built on the  
first null vector appearing at level $|m|+\nicefrac{1}{2}$, with 
$(-)^F=-1$. By induction we find the 
$(-)^{j-m}$ factor in eqn.~(\ref{ishicross}).}
\begin{equation}
\llangle \eta=-1;\, j\,m\, m|  e^{-\pi t H_\textsc{c}}
| j\, m \, m ;\, \eta=+1\rrangle =
(-)^{j-m}\, \frac{q^{\tfrac{j^2}{2}}+q^{\tfrac{(j+1)^2}{2}}}{\eta (it)}
\left( \frac{\vartheta_4 (it)}{\eta (it)}\right)^{\nicefrac{1}{2}}\, .
\label{ishicross}
\end{equation}
Then the annulus amplitude~(\ref{crossannulus}) reads
\begin{align}
\mathcal{A}_{-+}&= \sum_{j=0}^{\infty} \sum_{m=-j}^j (-)^m 
\llangle -1;\, j\,m\, m|  e^{-\pi t H_\textsc{c}}| j\, m \, m ;\, +1\rrangle
\nonumber\\
&= \sum_{j=0}^{\infty} (-)^j (2j+1) 
\frac{q^{\tfrac{j^2}{2}}+q^{\tfrac{(j+1)^2}{2}}}{\eta (it)}
\left( \frac{\vartheta_4 (it)}{\eta (it)}\right)^{\nicefrac{1}{2}}=
\left( \frac{\vartheta_4 (it)}{\eta (it)}\right)^{\nicefrac{3}{2}}
\nonumber \\
& = \left( \frac{\vartheta_2 (-\nicefrac{1}{it})}{
\eta(-\nicefrac{1}{it})}\right)^{\nicefrac{3}{2}} \ .
\end{align}
Therefore we find that in the open string channel (last line) the
annulus  amplitude between the brane 
$g=\mathbb{I}$ with $\eta=-1$ and the brane 
$g=i\sigma_3$ with $\eta=+1$ leads to the 
character for an {\it untwisted} 
$j=\nicefrac{1}{2}$ representation of the affine
$\widehat{\mathfrak{su}}(2)$  algebra  at level two. (From the point
of view of the free fermions, the $j=\nicefrac{1}{2}$ representation
describes the \textsc{r} sector, which should indeed appear between
branes of opposite $\eta$.)
This shows that these two Dirichlet branes sit at the {\it same} fixed
point of the super-orbifold. Computing instead the amplitude for two
$g=\mathbb{I}$ branes with $\eta=+1$ and $\eta=-1$, one finds a
twisted  $j=\nicefrac{1}{2}$ character, showing that these two
Dirichlet branes do not sit at the same fixed point. The important
conclusion of this computation is that the roles of the two fixed
points $y=0$ and $y=\nicefrac{\pi}{2}$ are exchanged when replacing
$\eta=+1$ by $\eta=-1$. 

\paragraph{Fractional branes correspondence for  $\bm{\eta=-1}$} 
With these preparations, we are now in the position to find the 
correspondence for the fractional branes of the circle and 
super-orbifold theories for $\eta=-1$. For $\eta=-1$, the only fixed
points of the circle theory correspond to diagonal group elements
and describe Dirichlet branes. They are resolved by odd momentum  
(\textsc{nsns} sector) and even momentum (\textsc{rr} sector) 
Dirichlet Ishibashi states. In the super-orbifold theory, these fixed points 
are mapped, via the identification (\ref{circorbmap}), to the group elements
(\ref{sof1}).

As explained above, the Dirichlet point $g=\mathbb{I}$ corresponds for
$\eta=-1$ to a Dirichlet brane sitting at the fixed point
$y=\nicefrac{\pi}{2}$.  
According to the correspondence it is resolved by twisted sector Dirichlet 
Ishibashi states from this fixed point, both in the \textsc{nsns} and
\textsc{rr} sectors.  
This is in perfect agreement with the analysis of the $\eta=+1$ case:
in the twisted \textsc{ns} sector, there are fermionic zero modes
satisfying~(\ref{diriground}). If the ground state at $y=\pi/2$ satisfies a
Neumann boundary condition to $\eta=+1$, it automatically satisfies
a Dirichlet boundary condition for $\eta=-1$, and thus the fixed point at 
$y=\pi/2$ supports a Dirichlet Ishibashi state in the twisted
\textsc{nsns} sector 
for $\eta=-1$. On the other hand, the twisted \textsc{rr} sector does
not have fermionic zero modes, and hence the existence of the
Dirichlet Ishibashi state  at $y=\pi/2$ for $\eta=+1$ implies also
that such a Dirichlet Ishibashi state  
exists for $\eta=-1$.

\begin{figure}[!ht]
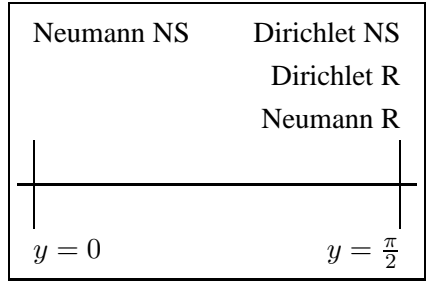

\centering
\fbox{\small \begin{tabular}{lcr}
Neumann NS & &  Dirichlet NS\\
&& Dirichlet R\\
&& Neumann R\\
\vline &&  \vline \\
\hline 
\vline && \vline \\
$y=0$ && $y=\tfrac{\pi}{2}$ 
\end{tabular}} 
\caption{\it Twisted sector branes for $\eta=-1$ in the super-orbifold
  theory. }   
\label{etaminusfig}
\end{figure}

As regards the `Neumann point' $g=i\sigma_1$ of the super-orbifold
theory, this is now resolved by a twisted \textsc{nsns} sector Neumann
Ishibashi state from  
the fixed point $y=0$, together with a twisted \textsc{rr} sector
Neumann Ishibashi state from the fixed point
$y=\nicefrac{\pi}{2}$. (The contribution of the two fixed 
points is therefore again asymmetric, as already before for
$\eta=+1$.) The contributions of the fixed points  
for $\eta=-1$  are summarised in fig.~\ref{etaminusfig}. 
The fact that all of this ties in very nicely together is convincing
evidence that the proposed one-to-one correspondence between the
branes of the two theories is indeed correct.  

Using these identifications one can now find linear relations between
the momentum Ishibashi states of the circle theory, and the fixed
point Ishibashi states of the super-orbifold theory.\footnote{
Strictly speaking we have so far only identified the twisted
\textsc{rr} sector Ishibashi states from the fixed point at
$y_0=\pi/2$; the identification for the Ishibashi states coming from 
the other twisted sector (at $y_0=0$) is however also uniquely fixed,
since these Ishibashi states must be `orthogonal' to the ones from
$y_0=\pi/2$, as the \textsc{lhs} of eqns.~(\ref{dic1},\ref{dic2}) in the 
bosonic theory.}
This will allow one to write explicit expression for the fractional
brane boundary states for generic group elements~(\ref{sof1})
(together with~(\ref{sof2}) for $\eta=+1$). The resulting formulae are
however not very illuminating, and since the structure is similar to
the bosonic case worked out in section~\ref{cequalone}, we shall not 
present them in detail here.

\section{Conclusions and open problems}

In this paper we have studied the complete boundary (D-brane) spectrum 
for multicritical points where the underlying closed string theory has
two equivalent, but superficially different, descriptions. In
particular we have considered the multicritical point of the $c=1$
bosonic theories where there is an equivalence between a circle theory
and an orbifold theory. We have also considered the multicritical
point of the $\hat{c}=1$, $\mathcal{N}=1$ superconformal
theories where the circle theory is equivalent to the so-called
super-orbifold theory. In each case, the two different descriptions
lead to a `canonical' spectrum of D-branes which looks a priori
different. We have checked that in both cases the full conformal
(resp.\ superconformal) D-brane spectrum of the two descriptions
however agrees. This suggests that the equivalence of these
multicritical theories also extends to the boundary  sector. 

In establishing this correspondence we have constructed a family
of `fractional' D-branes of the orbifold (and super-orbifold) theory
that interpolates between a standard fractional Dirichlet and a
standard fractional Neumann brane. Finally we have seen that the
natural starting point for the description of all superconformal 
$\mathcal{N}=1$ branes at $\hat{c}=1$ is the superaffine theory
(rather than the circle theory). This is the theory that has an affine
$\hat{\rm so}(3)_1$ symmetry.

It remains an open problem to describe similar phenomena for conformal
field theories with higher central charge. However, the complete moduli space of all conformal 
D-branes for such conformal field theories is not known; the description of all of 
these branes is likely to be a complicated problem since the number of
conformal Ishibashi states grows out of control for central charges
bigger than one.\footnote{Strictly speaking the relevant quantity is
the {\it effective} central charge.}
The known constructions of one theory will then typically correspond  
to branes that are unfamiliar from the other point of view; it is
therefore not obvious how one should compare the brane spectra of
these theories.

\subsection*{Acknowledgments}
We thank T.~Banks, I.~Brunner, M.~Douglas, S.~Elitzur, D.~Friedan, E.~Kiritsis, Y.~Oz, B.~Pioline, O.~Schlotterer, V.~Schomerus, N.~Seiberg and S.~Shenker for useful discussions.  
The work of MRG is supported in parts by the Swiss National Science Foundation and the Marie Curie
network `Constituents, Fundamental Forces and Symmetries of the
Universe'  (MRTN-CT-2004-005104). The work of E.R. was partially supported by the European Union Marie Curie
 RTN network under contract  MRTN-CT-2004-512194, the American-Israel Bi-National Science Foundation, 
the Israel Science Foundation, The Einstein Center in the Hebrew University, and by a grant of DIP (H.52). 
E.R. and D.I. acknowledge support from the France-Israel collaboration grant.

\appendix

\section{Conventions and definitions}
\label{appconv}
\paragraph{Pauli matrices}
The Pauli matrices are defined as 
\begin{equation}
\sigma_1 = \left( \begin{array}{cc} 0& 1 \\ 1 & 0 \end{array}\right)
\ , \qquad 
\sigma_2 = \left( \begin{array}{cc} 0& -i \\ i & 0 \end{array}\right)
\ , \qquad 
\sigma_3 = \left( \begin{array}{cc} 1& 0 \\ 0 & -1 \end{array}\right)\ .
\end{equation}
\paragraph{SU(2) matrix elements}
The SU(2) matrix elements in the representation of spin $j$ read
\begin{multline}
D^j_{m,\bar m}(g)  
=\!\!\! \sum_{\ell =\max(0,\bar m-m)}^{\min(j-m,j+\bar m)}
\frac{\sqrt{ (j+m)!\, (j-m)!\, (j+\bar m)!\, (j-\bar m)!}}
{(j-m-\ell)!\, (j+\bar m-\ell)!\, \ell!\, (m-\bar m+\ell)!}\ \times \\ 
\times \  
a^{j+\bar m-\ell} (a^*)^{j-m-\ell} b^{m-\bar m+\ell} (-b^*)^{\ell} \ ,
\label{matel}
\end{multline}
where the SU(2) group element $g$ is written as 
$g= \oaop{a\ \ b}{-b^\star\ a^\star}$ 
with $aa^\star +b b^\star =1$. 

\paragraph{Theta-functions}
The theta-functions are defined as ($q=\exp 2i\pi \tau$ and 
$z=\exp 2i\pi \nu$)
\begin{subequations}
\begin{align}
\vartheta_1 (\nu,\tau)& = \vartheta \oao{1}{1} (\nu;\tau) = 
i \sum_{n \in \mathbb{Z}} (-)^{bn} 
q^{\frac{1}{2} (n+\tfrac{1}{2})^2} z^{n+\tfrac{1}{2}}
\\
& = 2 q^{\frac{1}{8}} \sin \pi \nu  \prod_{m=1}^\infty 
(1-q^m)(1-zq^m)(1-z^{-1} q^m)\nonumber \\
\vartheta_2 (\nu,\tau)& = \vartheta \oao{1}{0} (\nu;\tau) = 
 \sum_{n \in \mathbb{Z}}  q^{\frac{1}{2} (n+\tfrac{1}{2})^2} z^{n+\tfrac{1}{2}}
\\
& =2 q^{\frac{1}{8}} \cos \pi \nu  \prod_{m=1}^\infty 
(1-q^m)(1+zq^m)(1+z^{-1} q^m)
\nonumber \\
\vartheta_3 (\nu,\tau)& = \vartheta \oao{0}{0} (\nu;\tau) = 
 \sum_{n \in \mathbb{Z}}  q^{\frac{1}{2} n^2}  z^n
=   \prod_{m=1}^\infty 
(1-q^m)(1+zq^{m-\frac{1}{2}})(1+z^{-1} q^{m-\frac{1}{2}})\\
\vartheta_4 (\nu,\tau)& = \vartheta \oao{0}{1} (\nu;\tau) = 
 \sum_{n \in \mathbb{Z}} (-)^{bn} q^{\frac{1}{2} n^2} z^n =
\prod_{m=1}^\infty 
(1-q^m)(1-zq^{m-\frac{1}{2}})(1-z^{-1} q^{m-\frac{1}{2}}) \ .
\end{align}
\end{subequations}
We use thereafter the shorthand notation 
$\vartheta_i (\tau):= \vartheta_i (0,\tau)$. The Dedekind eta-function
is 
\begin{equation}
\eta (\tau)= q^{\frac{1}{24}} \prod_{m=1}^\infty (1-q^m)\ .
\end{equation}
Some useful identities are
\begin{equation}
\vartheta_2 (\tau)\vartheta_3 (\tau)\vartheta_4 (\tau) 
= 2\eta^3 \ , \qquad 
\vartheta_4 (2\tau) = \sqrt{\vartheta_3(\tau) \vartheta_4 (\tau)} \ .
\end{equation}
Finally the modular transformations read
\begin{subequations}
\begin{align}
\vartheta \oao{a}{b} (\nu,\tau+1) &= e^{-\frac{i\pi}{4} a(a-2)} 
\vartheta \oao{a}{a+b-1} (\nu,\tau)\\
\vartheta \oao{a}{b} (\nicefrac{\nu}{\tau},-\nicefrac{1}{\tau}) & 
= \sqrt{-i\tau} 
e^{\frac{i\pi ab}{2}+ \frac{i\pi \nu^2}{\tau}} 
\vartheta \oao{-b}{a} (\nu,\tau) \ . 
\end{align}
\end{subequations}

\end{document}